\makeatletter
\def\input@path{{article/}{article/styles/}}
\makeatother
\documentclass{article/optica-article}

\journal{opticajournal} 

\articletype{Research Article}

\usepackage{lineno}
\usepackage{xcolor}
\usepackage[normalem]{ulem}

\usepackage[version=4]{mhchem}

\begin{document}

\newcommand{\red}[1]{\textcolor{red}{#1}}
\title{Non-invasive super-resolution imaging through scattering media using highly nonlinear labels}

\author{Pawel~Szczypkowski\authormark{1,\dag}, Adrian~Makowski\authormark{1,2,\dag}, Wojciech~Zwoliński\authormark{1}, Andrzej~Kozłowski\authormark{1}, Katarzyna~Prorok\authormark{3}, Piotr~Wasylczyk\authormark{1}, Artur~Bednarkiewicz\authormark{3} and Radek~Lapkiewicz\authormark{1,*}}

\address{\authormark{1}Faculty of Physics, University of Warsaw, Warsaw, Poland\\
\authormark{2}Laboratoire Kastler Brossel, ENS--PSL Université, CNRS, Sorbonne Université, Collège de France, Paris, France\\
\authormark{3}Institute of Low Temperature and Structure Research, Polish Academy of Sciences, Wroclaw, Poland\\
\authormark{\dag}The authors contributed equally to this work}

\email{\authormark{*}Radek.Lapkiewicz@fuw.edu.pl}

\begin{abstract*} 
While scattered light conveys most of the information we perceive, scattering may also distort that information before it reaches our detectors. The problem is acute in many applications, such as in high-resolution microscopy of biological tissue, where scattering degrades both resolution and signal-to-noise ratio. Here, for the first time, we demonstrate that combining two intrinsic properties of scattered light: speckle statistics and the memory effect, with highly non-linear optical response yields, rather surprisingly, super-resolution, low-background, non-invasive imaging of objects completely hidden behind a strongly scattering, opaque layers. Crucially, our technique of Nonlinear Imaging with Speckle Excitation (NISE)  does not resort to wavefront shaping, adaptive optics, complicated optical setups, or iterative image reconstruction algorithms. Because the strategy relies solely on the properties of scattered light and high-order nonlinear response of the luminescent labels, it can be applied to any speckle-forming propagation, from biological tissue to multicore fibers, combined with any type of phenomenon that exhibits a sufficiently high order nonlinearity.
\end{abstract*}

\section*{Introduction}

Light scattering is a fundamental challenge in modern optics -- it limits the performance of early, specific medical diagnostics, targeted photodynamic therapies, or examination of complex biological systems, such as brain neurons in vivo. In fluorescence microscopy of neuronal activity, scattering severely degrades both resolution and the signal-to-noise ratio, obscuring fine structural details and limiting imaging to superficial layers of specimens. The standard technique of deep imaging - multiphoton microscopy~\cite{Denk1990,Helmchen2005,Horton2013} -- suffers from the unfavorable exponential decay of ballistic light intensity with depth: scattering degrades resolution and ultimately limits the achievable imaging depth. Addressing that issue, Vellekoop and Mosk demonstrated that scattering can be compensated by wavefront optimization~\cite{Vellekoop2007,Vellekoop2010} and scattered light can be shaped into a usable excitation focus. Subsequent work revealed that scattering can be described with linear operators acting on the input optical field~\cite{Popoff2010a,Popoff2010b}, enabling wavefront control of light propagation through complex media. Building on these insights, researchers explored the phenomenon of memory effect~\cite{Freund1988,Osnabrugge2017} to image samples hidden behind scattering layers~\cite{Yoon2020,Gigan2022,Bertolotti2022}, either with wavefront shaping~\cite{Horstmeyer2015,Katz2012,Katz2014a,Lai2015,Papadopoulos2017}, or computationally~\cite{Bertolotti2012,Katz2014b,Boniface2020,Zhu2022,Weinberg2024}. While these and other approaches, from computational holography~\cite{Haim2025} to phase conjugation~\cite{Yaqoob2008}, have greatly advanced the field, they remain diffraction-limited and require complex hardware and/or highly demanding reconstruction algorithms.

Here, we demonstrate that super-resolution imaging through strongly scattering media can be achieved without the need for time-consuming wavefront optimization or computational overhead. Our method of Nonlinear Imaging with Speckle Excitation (NISE) relies solely on basic principles of light propagation, requires minimal hardware–software complexity, and exhibits favorable power scaling with depth, compared to the approaches using ballistic light. By exploiting the diffraction-limited size of speckle grains formed after propagating in scattering media, and labels with high luminescence nonlinearity (e.g. photon avalanching particles~\cite{Lee2021,Bednarkiewicz2019,Duh2020}), NISE opens new opportunities in optical sensing and super-resolution imaging in biology, medicine, and material science. Being fundamentally different from the previous approaches tackling light scattering, our method also has the potential to create a new domain of research in imaging with scattered light.

\section*{Principle}
Conventional nonlinear techniques used for imaging in or through scattering media, such as two- or three-photon microscopy, rely on ballistic photons to achieve localized fluorescence emission. However, beyond a depth of a few transport mean free paths, or when imaging through highly scattering layers, the ballistic light intensity becomes prohibitively low for the multiphoton approaches~\cite{Denk1990,Helmchen2005,Horton2013}.

\begin{figure}[htbp]
	 \captionsetup{font={footnotesize,stretch=0.95}}
    \centering
    \includegraphics[width=1\textwidth]{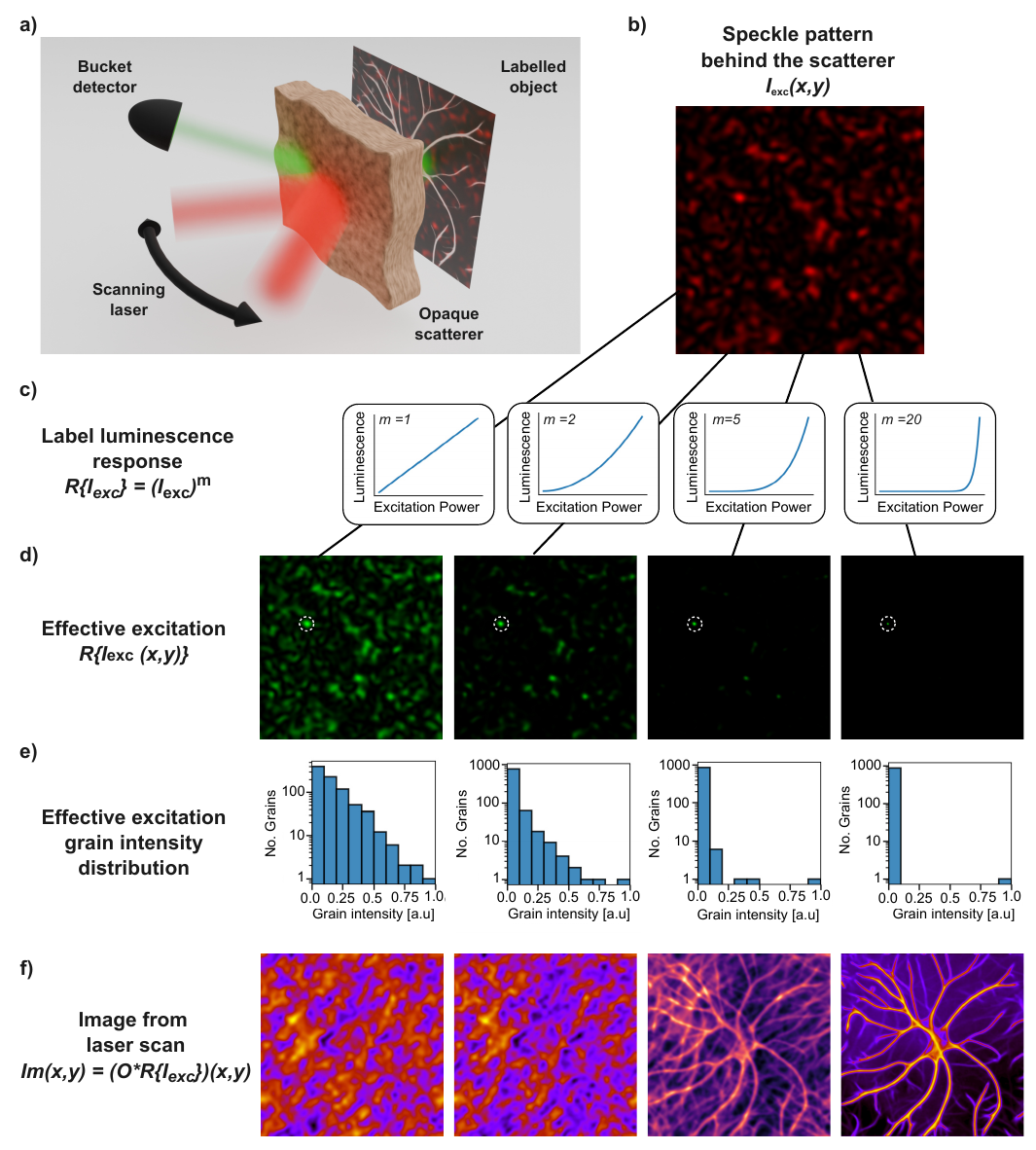} 
	\caption{\textbf{Schematics and numerical simulations illustrating the principle of Nonlinear Imaging with Speckle Excitation (NISE).} 
    Labeled object $O$ (based on an astrocyte image) is completely hidden behind an opaque scattering layer and is illuminated with a laser light that excites luminescent labels (a). After propagation through the scatterer, the beam reaches the object as speckles (b). The effective excitation pattern depends on the nonlinearity of the label response: $R\{I_{\mathrm{exc}}\} = (I_{\mathrm{exc}})^{m}$, where $R\{\cdot\}$ is the luminescence response function, and $I_{\mathrm{exc}}$ is the excitation light intensity. For linear response (left column: $m=1$), the effective excitation is a dense, grainy pattern, while for increasing nonlinearity (middle columns: $m=2$, $m=5$), it becomes more sparse, until it consists of a single sub-diffraction excitation spot originating from the brightest speckle (right column: $m=20$, marked with white circle), while all the others are suppressed (d). The histograms of the effective excitation speckle grain brightness (in normalized arbitrary units) in the log scale (e). Images recorded by raster scanning of the speckle pattern across the object are given by the convolution of the object and the effective excitation: $I_{m}(x,y) = (O * R\{I_{\mathrm{exc}}\})(x,y)$ (f).}
	\label{fig:fig1} 
\end{figure}

NISE allows for direct imaging of samples completely hidden behind opaque (i.e. transmitting no ballistic light) scattering layers (Fig.~\ref{fig:fig1}a). When an excitation laser beam propagates through the scatterer, it forms a speckle intensity pattern $I_{\mathrm{exc}}(x,y)$ (Fig.~\ref{fig:fig1}b). The speckle pattern consists of diffraction-limited grains formed by interference of many contributions from small features in the scatterer. While scattering causes the overall illumination pattern to broaden, the spatial extent of individual speckle grains does not increase during propagation in the scattering medium (see Supplement~1 Section~2.C). This speckle grain size is given by the diffraction limit $\lambda/(2\mathrm{NA})$~\cite{Goodman2020}, where $\lambda$ is the wavelength of scattered light and NA is the effective numerical aperture~\cite{Vellekoop2010}. If the sample is stained with labels that luminesce in response to the speckle excitation intensity, the spatial distribution of the emitted luminescence intensity $I_{\mathrm{lum}}(x,y)$ can be described as:
\begin{equation}
I_{\mathrm{lum}}(x,y) \;=\; R\{I_{\mathrm{exc}}(x,y)\}\,\cdot\, O(x,y),
\end{equation}
Where $O(x,y)$ is the (unknown) spatial distribution of the luminescent labels, and $R\{\cdot\}$ describes the optical (luminescent) response of the labels (Fig.~\ref{fig:fig1}c). For the linear optical response of the labels (i.e.\ $R\{I\}\propto I$) the luminescence intensity at a given point is proportional to the local illumination intensity, resulting in a grainy effective excitation $R\{I_{\mathrm{exc}}(x,y)\}$ (Fig.~\ref{fig:fig1}d, left). Due to the statistical nature of speckles, the brightness of the grains of the effective excitation follows the negative exponential distribution~\cite{Goodman2020} (Fig.~\ref{fig:fig1}e, left). With the linear response of the labels and the opaque scatterer, raster scanning of the laser beam inevitably produces a noisy image (Fig.~\ref{fig:fig1}f, left).

The situation changes when the response of the labels is nonlinear. Surprisingly, for sufficiently high nonlinearity of the optical response ($R\{I_{\mathrm{exc}}\}\propto (I_{\mathrm{exc}})^m$), the effective excitation becomes localized to a single, sub-diffraction spot. The middle and right columns in Fig.~\ref{fig:fig1} present the situation for $m=2$, $m=5$, and $m=20$. The higher the nonlinearity, the sparser the effective excitation: the brightest speckle stands out from the others, as it is unlikely for comparably bright speckles to appear (see Supplement~1 Section~1. for detailed statistical analysis). With increasing nonlinearity $m$, the contribution of the brightest speckle to the effective excitation becomes increasingly pronounced, while contributions from speckle grains with lower intensities get suppressed. Fundamental speckle properties ensure that the effective excitation becomes localized to a single spot that is also of sub-diffraction size due to the nonlinearity (See Supplement~1 Sections~3. and~4.). Thus, remarkably, illumination with a random speckle field produces luminescence equivalent to the one produced by illumination with a tightly focused beam. Additionally, the sub-diffraction size of the excitation is preserved along the propagation depth inside the scattering sample.

To record an image of the object, the laser beam is scanned and the luminescence signal $I_{\mathrm{lum}}(x,y)$ (which also undergoes scattering) is collected with a bucket detector, placed at the illumination side of the scattering medium (i.e. in the epi-/non-invasive detection configuration, Fig.~\ref{fig:fig1}a). When the laser beam is tilted (scanned), within the memory effect range~\cite{Freund1988,Osnabrugge2017,Bertolotti2012}, the excitation speckle pattern is shifted by a distance proportional to the tilt angle $\phi_x$, $\phi_y$ and to the distance $d$ between the sample and the scattering layer. For each tilt angle, the recorded luminescence is proportional to:
\begin{equation}
I_m(\phi_x,\phi_y) \;=\; \iint R\!\left\{ I_{\mathrm{exc}}\!\left(x-\phi_x d,\, y-\phi_y d\right) \right\}\, O(x,y)\, dx\,dy.
\end{equation}
The results of such scans with the increasing nonlinearity of the labels are shown in Fig.~\ref{fig:fig1}f. With the effective excitation $R\{I_{\mathrm{exc}}(x,y)\}$ dominated by a single speckle grain (Fig.~\ref{fig:fig1}d right), this integral reduces to sampling the label distribution within the object at the location of the brightest speckle. This leads directly to the recording of an image of the object hidden behind a scatterer with sub-diffraction resolution, given by $\lambda/(2\mathrm{NA}\sqrt{m})$ (See Fig.~\ref{fig:fig1}f right and Supplement~1 Section~4.).

\section*{Experiment}

For experimental demonstration of NISE, we use a custom laser scanning microscope with a continuous-wave \(1064\,\mathrm{nm}\) laser excitation and detection of the \(800\,\mathrm{nm}\) up-conversion emission from a sample of avalanching particles \cite{Lee2021}. The non-invasive configuration uses the epi detection and the objective with a long working distance of 8.2 mm and the NA of 0.45 (Nikon CFI S Plan Fluor ELWD 20XC). The excitation path of the microscope includes a scanning module based on two optically conjugated galvo mirrors (2x Thorlabs: GVS011) and the beam expansion optics. The galvo mirrors are imaged onto the back focal plane of the objective, with the beam filling the back aperture of the objective completely. The 800 nm luminescence emitted from the sample is separated from the excitation light with a dichroic mirror (Thorlabs: DMS900R) and a set of spectral filters (2X FESH0900, and 2X NF1064-44), and recorded with a silicon photon multiplier (Thorlabs: PDA41), used as a bucket detector. We use a scan pixel dwell time of 5 ms and average over 1000 voltage collected samples using a DAQ card during the dwell time. The raster scanning with a sawtooth pattern required no postprocessing to obtain an image. The detailed schematic of the experimental setup with extended description can be found in the Supplement~1 Section~5.

For NISE experiments, a scattering layer is introduced in the excitation path between the sample and the microscope objective, typically 3–4 mm from the sample. For our experiments two types of the scattering layers were used: either a 0.5$^{\circ}$ holographic diffuser (Edmund Optics: \#47-988) transmitting negligible amount of ballistic light (used in experiments in Fig.~\ref{fig:fig2} and ~\ref{fig:fig3}), or fresh calf brain tissue slices (480 and 720 $\mu$m thick) -- Fig.~\ref{fig:fig4}.

The experimental procedure is as follows: We take the ground truth image of the sample with a focused beam (no scatterer in the beam path). Next, we put the scatterer in front of the sample (around 3-4 mm from the sample). We change the position of the sample along the optical axis, effectively finding a "best-effort" focus based on the luminescence response of the sample. Subsequently, we perform scanning with the speckle pattern across the sample with avalanching micro-crystals or avalanching nanoparticles (a detailed procedure of sample preparation is described in Supplement~1 Section~6.). The 800 nm luminescence emitted from the sample travels back through the scatterer, is separated from the residual excitation light with a dichroic mirror and filters, and is focused on the silicon photon multiplier with an aspheric lens.

After taking the scan, we inspect the speckle pattern. Importantly, this step is not required for NISE and serves for experiment inspection only.  The detailed procedure of speckle pattern measurement is described in Supplement~1 Section~5.C.

Although, in principle, there is no need for finding an exact focus in NISE, we find a rough, "best-effort focus" for three reasons. Firstly, the excitation pattern has the smallest spatial extent in the focus of the objective, which reduces the overall excitation power needed for imaging (which is crucial while imaging through biological samples, such as the brain slice). Secondly, the small spatial extent of speckle pattern increases the probabilistic chance that the brightest speckle has considerably higher peak intensity than the others (see the Supplement~1 Section~1.E). Lastly, the speckle pattern around the geometrical focus of the objective shifts by the same amount as the focused beam during scanning and thus no scaling needs to be applied to compare the ground truth and NISE images.

\section*{Results}
\begin{figure}[htbp]
	\centering
	\includegraphics[width=1.0\textwidth]{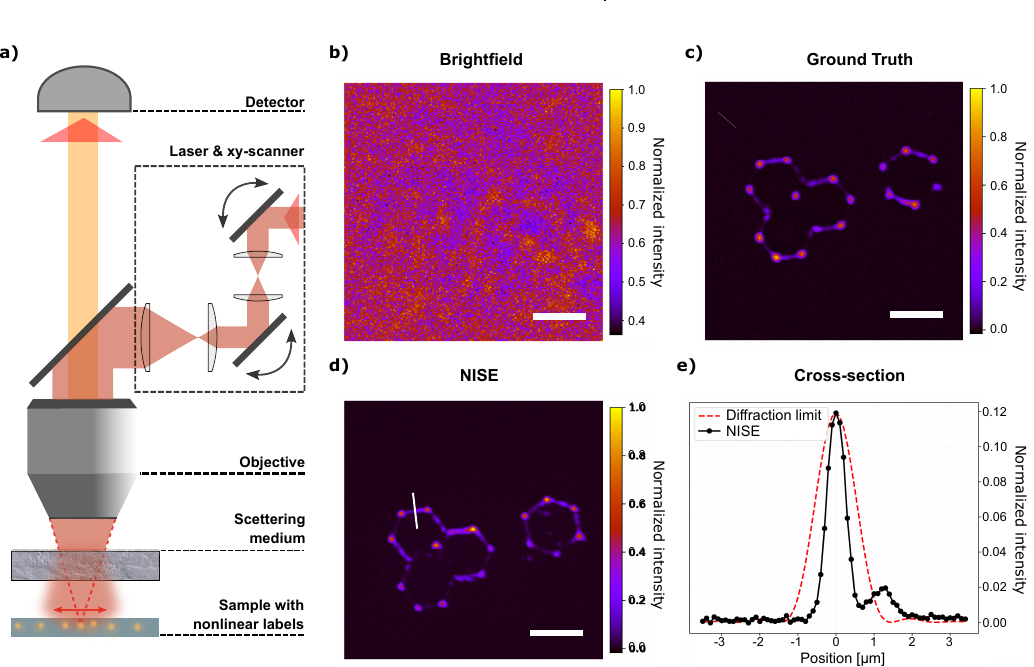}

	\caption{\textbf{Non-invasive super-resolution imaging through an opaque scattering layer.} 
    (a) Experimental setup: a laser scanning optical microscope with a \(1064\,\mathrm{nm}\) continuous wave laser excitation. After propagating through a scattering layer, the laser beam excites the sample—avalanching hexagonal microcrystals, and their luminescence is detected in the epi configuration by a bucket detector. (b) Brightfield image of the sample recorded through the scatterer. (c) Ground truth—laser scanning microscope image of the sample recorded without the scatterer. (d) NISE image of the sample (photon avalanching hexagonal microcrystal) recorded through the same scatterer.  (e) Cross-section of the edge of the microcrystal along a white line marked in (d) (black points), compared with the diffraction limited PSF (red dashed line). Scale bars in (b)--(d): \(10\,\mu\mathrm{m}\).}
	\label{fig:fig2} 
\end{figure}

Utilizing speckle excitation, we first imaged a sample of NaLuF\(_4\) avalanching microparticles. The microparticles, doped with \(8\,\mathrm{mol}\,\%\) Tm\(^{3+}\) ions, exhibit high nonlinearity with \(m \simeq 10\), photon-avalanching thresholds of tens of \(\mathrm{kW\,cm^{-2}}\) photoexcitation intensity, excellent photostability, and bright \(800\,\mathrm{nm}\) emission under \(1064\,\mathrm{nm}\) photoexcitation (see Supplement~1 Section~6.). While imaged through a scattering layer in a standard brightfield modality with trans-illumination, the microparticles are not visible due to strong light scattering (Fig.~\ref{fig:fig2}b). However, in NISE, due to their highly nonlinear response to the excitation light intensity, the effective excitation is confined to a single speckle grain, with negligible contribution of luminescence induced by other speckles. The resulting image of the sample presented in Fig.~\ref{fig:fig2}d shows excellent agreement with the one obtained under scattering-free conditions (Fig.~\ref{fig:fig2}c), demonstrating the efficacy of our method when no ballistic light reaches the sample.

Additionally, our approach intrinsically enables super-resolution imaging: the speckle grain size is diffraction-limited, and the non-linear response of the avalanching particles narrows the excitation point spread function (PSF). Using an objective with a numerical aperture (\(\mathrm{NA}\)) of \(0.45\) and an excitation wavelength of \(1064\,\mathrm{nm}\), we achieve a full-width at half-maximum transverse resolution of \(540\,\mathrm{nm}\) -- significantly below the theoretical diffraction limit of \(1.2\,\mu\mathrm{m}\). A quantitative comparison between the diffraction-limited PSF (Airy function computed for \(1064\,\mathrm{nm}\) wavelength and \(\mathrm{NA}=0.45\)) and the measured edge profile of a microcrystal from the image in Fig.~\ref{fig:fig2}d is presented in Fig.~\ref{fig:fig2}e.

\begin{figure} [htbp]
	\centering
	\includegraphics[width=0.99\textwidth]{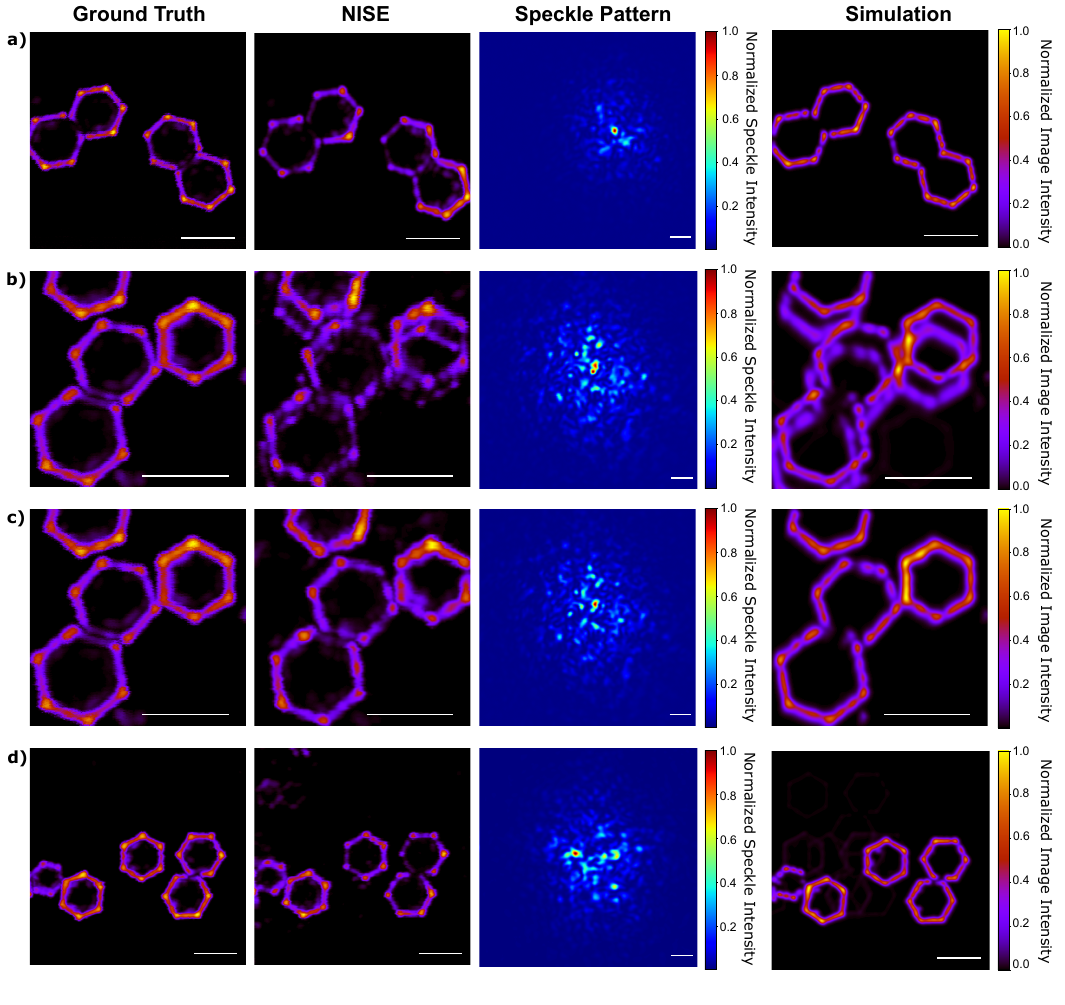}
	\caption{\textbf{Nonlinear Imaging with Speckle Excitation.} Comparison of NISE for different speckle excitation patterns. The first column presents laser scanning microscope images of the sample of avalanching microcrystals recorded without the scatterer (ground truth). The second column presents the experimental result of NISE. The third column shows the speckle pattern measured in the sample plane with scanning over a pinhole (1~$\mu$m diameter in (a-c) and 2~$\mu$m in (d)) . The last column shows the simulated result of NISE assuming the nonlinearity of m=10. In row (a) the speckle pattern constitutes mostly of a single speckle grain and this results in perfect resemblance of the NISE image with the ground truth image. In row (b) the speckle pattern has two grains with similar brightness and the resulting NISE image contains a double image, also clearly visible in the simulation. In row (c) the same region of interest was excited with speckle pattern in slightly different position along the optical axis. The speckle pattern id (c) differs slightly from (b) but now there is only one distinct bright speckle which result in perfect resemblence of the ground truth with the NISE image. Row (d) shows the NISE result for the speckle pattern with the brightest speckle offset from the center of the speckle pattern. Scale bars are 10~$\mu$m.}
	\label{fig:fig3} 
\end{figure}

\begin{figure} [htbp]
	\centering
	\includegraphics[width=0.99\textwidth]{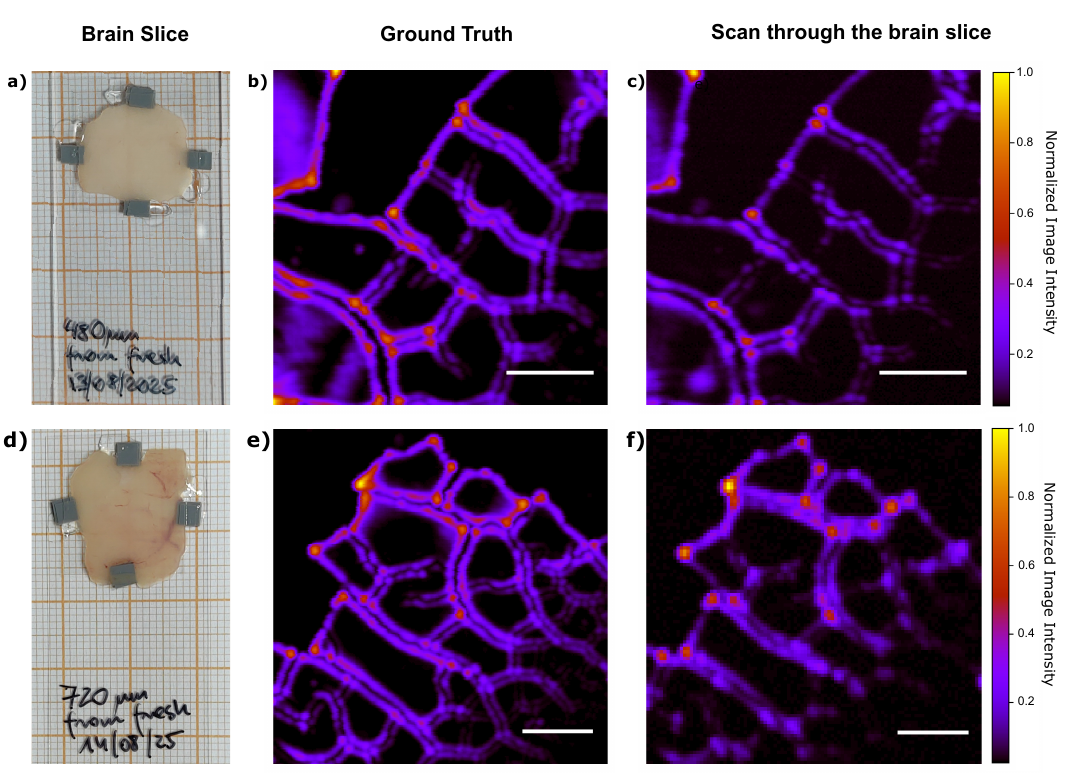}

	\caption{\textbf{Nonlinear Imaging through  thick brain slices.} 
    Left column (a and d) shows brain slices on a millimeter scale paper background. Image (c) presents NISE imaging of clusters of avalanching nanoparticles through a 480~$\mu$m brain slice. Image (f) is NISE imaging of the same sample through a 720~$\mu$m brain slice. The excitation power behind the objective was approximately 0.5 W for scan through the 480~$\mu$m brain slice and 0.8 W for the  720~$\mu$m slice. For imaging through the thicker brain slice, the pixel size was increased to 500 nm to reach the optimum pixel sampling and reduce heating and the acquisition time. Middle column shows the ground-truth laser scanning microscope images of the sample recorded without the scatterer. Scale bars are \(10\,\mu\mathrm{m}\).}
	\label{fig:fig4} 
\end{figure}

Figure \ref{fig:fig3} presents NISE images recorded for different (random) speckle patterns: one with a single brightest spot (row a), one with two comparably bright brightest spots (row b), one similar to (b), but with a slightly different speckle realization (row c), and one where the brightest speckle is well off-centered (row d). With the situation in row (a) NISE provides the high quality imaging, while the large nonlinearity increases the resolution below the diffraction limit. The excitation speckle pattern consists mostly of single distinct grain and such situation could also resemble a ballistic light focus with substantial intensity of scattered light around this focus. In (b), the two speckles of comparable intensity contribute to the "ghost" image, but a slight shift along the z axis results in a different speckle realization (row c) with one, single brightest speckle, and a clear image is restored. The images in row (d) prove that NISE indeed uses the brightest speckle, and not the ballistic light. In this case the brightest speckle grain is situated far from the speckle pattern center. The experimental images of NISE are compared with the ground truth images (focused beam scanning microscope with no scatterer) and the images from simulations. Each simulation is the convolution of the deconvolved ground-truth image with the effective excitation centered on the brightest speckle. The effective excitation is calculated by raising the measured speckle intensity pattern to the 10th power, corresponding to the measured nonlinearity order of the microcrystals. The code used to generate the simulated images and analyze the experimental data is available as Code~1 (Ref.~\cite{github_Szczypkowski2026}). For a relatively low nonlinearity there is still a good chance to have other speckles to be bright enough to contribute to the NISE image, as visible in row (b). The probability of such situation decreases with the increasing spatial extent of the speckle pattern and with increasing nonlinearity (see Supplement~1 Section~1.).

We also demonstrate that our method is effective for imaging through thick, strongly scattering biological tissues. A slice of fresh calf brain (480~$\mu$m in Fig.~\ref{fig:fig4}a  and 720~$\mu$m in d ) was inserted behind the objective, around \(3\,\mathrm{mm}\) from the sample. This time, the sample contained hexagonal \(\beta\)-NaYF\(_4\) \(20\,\mathrm{nm}\) diameter nanoparticles doped with \(8\,\mathrm{mol}\,\%\) Tm\(^{3+}\) ions, with a lower avalanching threshold than the NaLuF\(_4\) microparticles and with the nonlinearity \(m \simeq 6\). These nanoparticles were drop-cast on a glass slide, and after drying, formed quasi-crystalline domains. The edges of these domains exhibit the lowest avalanching threshold (see Supplement~1 Section~7.) and, as such, are the brightest in the scanning microscope image (Fig.~\ref{fig:fig4}b and e). The scans through the brain slices (Fig.~\ref{fig:fig4}c and f) match the result of a scan obtained with a focused beam without the scatterer placed in front of the sample. The examination of the speckle pattern behind the brain slice is not accurate as the tissue changes over time and so does the speckle pattern. We expect the imaging to mostly utilize the ballistic component and look similar to the speckle pattern in Fig.~\ref{fig:fig3}a. 

\clearpage

\section*{Discussion}
We introduced Nonlinear Imaging with Speckle Excitation (NISE) -- a fundamentally new way to extract (for the first time, to our knowledge) sub-diffraction information from behind opaque scattering layers by exploiting statistical properties of speckles combined with high optical nonlinearity of luminescent labels. The nonlinearity of the labels can originate from various physical processes, as our framework is general and does not rely on a particular characteristic of the nonlinear response. Introducing high optical nonlinearity as a third strategy of imaging in scattering media, alongside wavefront shaping and computational methods, opens a largely unexplored domain of research in pure and applied optics.

Due to its robustness and no need for complex instrumentation or reconstruction algorithms, NISE is a promising departure point for hybrid approaches, combining other hardware and software methods. For instance, computational reconstruction~\cite{Bertolotti2012,Katz2014b,Boniface2020,Zhu2022,Weinberg2024,Haim2025} may improve its performance when several speckles have comparable brightness. At high nonlinearities, even small differences in speckle brightness are strongly amplified, making it more likely that most of the luminescence originates from a single dominant speckle. At lower nonlinearities, several speckles may contribute appreciably to the image, in which case computational reconstruction and data-analysis methods could be used to recover the underlying object structure.

Most computational and wavefront shaping methods are fundamentally limited by the range of the memory effect, which decreases rapidly with depth. The situation is different for NISE where, beyond the memory-effect range, the speckle pattern changes, but retains the same statistics, allowing a new dominant speckle to emerge and enabling patch-wise imaging of extended structures. By combining this strategy with correlation techniques, the accessible field of view may be extended.

In many cases of e.g. bio-imaging the ballistic component remains more intense than the scattered component even for relatively large penetration depths. Scanning with ballistic light has the advantage of not being restricted by the memory effect. Even if the ballistic component is similar in intensity to the scattered component, imaging is still possible. The perhaps less obvious feature of NISE is that in some cases it is favorable to image with speckles, rather than with a focused beam. I. e. it could be beneficial to defocus from the highly aberrated ballistic focus to image with the brightest speckle. Speckle grains remain diffraction limited, regardless of the depth, while the ballistic focus deteriorates due to aberrations, and due to the decrease of the effective NA with depth~\cite{Yoon2022}. In contrast, the resolution of the image obtained with scanning the speckles is independent of the depth, and still increases by a factor of \(1/\sqrt{m}\) relative to the diffraction limit. Even though propagation through scattering media can introduce aberrations, speckles can be thought of an extreme beam distortion, which results in the method being insensitive to additional wavefront distortions.

The scaling of laser power, required for NISE, with depth is also favorable compared to methods based on ballistic light. In scattering media, such as tissues, ballistic light intensity in the focus decays exponentially with depth. In contrast, the intensity of the brightest speckle scales approximately as an inverse of the depth (see Supplement~1 Section~3.).

On the other hand, NISE inevitably requires high excitation laser powers, since it uses light from the brightest speckle for excitation, and only a small fraction of the power contributes to the image creation. While avalanching particles used in our experiments exhibit the required steep nonlinearities and are excitable at biocompatible and deeply penetrating \(1064\,\mathrm{nm}\) wavelength, they demand relatively high (\(10\text{--}100~\mathrm{kW\,cm^{-2}}\)) excitation intensities at the sample. Laser powers required to reach the photon avalanche threshold and the steep nonlinear response may not be compatible with, e.g., live biological imaging. Lowering the avalanche threshold will be important for reducing excitation intensities to more biocompatible levels. Parallel efforts in developing a toolbox of labels that combine strong nonlinear responses at other excitation and emission wavelengths will further broaden the applicability of NISE. Specific biological labeling with the avalanching particles is already available using protocols that had been developed for upconverting nanoparticles~\cite{Hlavacek2022}. Further advances in nanoparticle design and the prospect of engineered emitters with step-like responses suggest that many new solutions in this area may soon be within reach~\cite{Szalkowski2025}.

In its present form, NISE can be combined with wavefront shaping, using the nonlinear signal as an optimization feedback~\cite{Katz2014a}, to reduce power demands, while adding minimal optimization overhead. Its generality makes it adaptable to a wide range of systems; for example, fiber bundles can be treated as scattering media~\cite{Porat2016}, opening a route towards endoscopic implementations. Moreover, nonlinear nanoparticles can serve as multifunctional probes, such as pH, FRET, or temperature sensors~\cite{bednarkiewicz2020}, enabling noninvasive sensing in scattering environments. In a broader context, exploring the interaction of speckles with high nonlinearities opens a new direction in optical science: beyond microscopy, the same principles could be applied to material processing, endoscopy, laser surgery, photodynamic therapy, optogenetics, or optical data processing.

\section*{}

\begin{backmatter}
\bmsection{Funding}
We acknowledge the support of the following funding agencies: Foundation for Polish Science (FIRST TEAM project FENG.02.02-IP.05-0253/23); National Science Centre (grants 2023/49/N/ST7/04195 and 2022/47/B/ST7/03465); National Centre for Research and Development QuantERA II projects: QM3 (QuantERAII/02/QM3/03/2024) and EXTRASENS (QuantERAII/02/EXTRASENS/02/2024); HORIZON EUROPE Marie Sklodowska-Curie Actions (FLORIN ID 101086142); and the French Government Scholarship for Ph.D. Cotutelle/Codirection. AB and KP acknowledge the support of the National Science Centre grant 2021/43/B/ST5/01244.

\bmsection{Acknowledgment}
We thank T. Stefaniuk for taking electron microscopy images of nanoparticles and S. Gigan for insightful discussions. The authors thank E.Bukowska and O.Bezkrovnyi for XRD and TEM measurements, respectively.

\bmsection{Disclosures}
Authors declare that they have no competing interests.

\bmsection{Data availability}
Data underlying the results presented in this paper are not publicly
available at this time but may be obtained from the authors upon
reasonable request. The code used to generate the simulations and analyze
the experimental data is publicly available as Code~1
(Ref.~\cite{github_Szczypkowski2026}).

\bmsection{Supplemental document}
See Supplement~1 for supporting content.

\end{backmatter}

\resetlinenumber[1]

\clearpage

\begingroup

\makeatletter
\@ifundefined{toclevel@section}{}{\def\toclevel@section{2}}
\@ifundefined{toclevel@subsection}{}{\def\toclevel@subsection{3}}
\@ifundefined{toclevel@subsubsection}{}{\def\toclevel@subsubsection{4}}
\@ifundefined{toclevel@paragraph}{}{\def\toclevel@paragraph{5}}
\@ifundefined{toclevel@subparagraph}{}{\def\toclevel@subparagraph{6}}

\@ifundefined{theHsection}{}{\renewcommand*{\theHsection}{supplement.\arabic{section}}}
\@ifundefined{theHsubsection}{}{\renewcommand*{\theHsubsection}{supplement.\arabic{section}.\arabic{subsection}}}
\@ifundefined{theHsubsubsection}{}{\renewcommand*{\theHsubsubsection}{supplement.\arabic{section}.\arabic{subsection}.\arabic{subsubsection}}}
\@ifundefined{theHfigure}{}{\renewcommand*{\theHfigure}{supplement.\arabic{figure}}}
\@ifundefined{theHtable}{}{\renewcommand*{\theHtable}{supplement.\arabic{table}}}
\@ifundefined{theHequation}{}{\renewcommand*{\theHequation}{supplement.\arabic{equation}}}
\makeatother

\makeatletter
\@ifundefined{phantomsection}{}{\phantomsection}
\@ifundefined{pdfbookmark}{}{\pdfbookmark[1]{Supplementary Information}{supplementary-information}}
\makeatother

\setcounter{section}{0}

\setcounter{subsection}{0}

\setcounter{subsubsection}{0}

\setcounter{figure}{0}

\setcounter{table}{0}

\setcounter{equation}{0}

\renewcommand{\thesection}{S\arabic{section}}

\renewcommand{\thesubsection}{S\arabic{section}.\arabic{subsection}}

\renewcommand{\thesubsubsection}{S\arabic{section}.\arabic{subsection}.\arabic{subsubsection}}

\renewcommand{\thefigure}{S\arabic{figure}}

\renewcommand{\thetable}{S\arabic{table}}

\renewcommand{\theequation}{S\arabic{equation}}

\begin{center}
  {\Large\bfseries Supplementary Information\par}
  \vspace{0.8em}
  {\large Non-invasive super-resolution imaging through scattering media using highly nonlinear labels\par}
\end{center}
\vspace{1em}

\section{Speckle grain intensity statistics}\label{sec:spk_stat}

In this section, we describe the statistical properties of speckles. 
The NISE technique relies on the amplification of intensity differences between speckles and is based on the presence of such variations. 
We investigate the probability of obtaining such favourable conditions for imaging (summarised in Sec.~\ref{sec:spk0}).
In Sec.~\ref{sec:spk_stat_1}, we present the general characteristics of speckles and introduce the procedure for distinguishing individual speckles. 
Sec.~\ref{sec:spk_1st} discusses the statistical distribution of the intensity of the brightest speckle. 
In Sec.~\ref{sec:spk_1st_2nd}, we extend this analysis to include both the brightest and the second brightest speckles. 
Finally, in Sec.~\ref{sec:spk_ratio} we derive the cumulative distribution function of the ratio $R$ between the intensity of the second brightest speckle $I_{2}$ and the brightest one $I_{1}$, and estimate the corresponding imaging probability.

\subsection{Statistical conditions for NISE}
\label{sec:spk0}
Although the exact speckle pattern cannot be predicted without detailed knowledge of the scatterer, speckle grains exhibit robust statistical properties: for a fixed scatterer and illumination, the average speckle grain size is well defined, intensities of grains in uncorrelated regions are independent, and the intensity distribution is well approximated by a negative exponential law~\cite{supp:Goodman2020}. Consequently, speckle brightness is not uniform, and there exists a brightest speckle.

Our method leverages the intensity gap between the brightest speckle grain with intensity $I_{(1)}$ and the others -- in particular, the second brightest speckle grain with intensity $I_{(2)}$. A nonlinear response of order $m$ amplifies this gap; the amplification depends on the ratio
\begin{equation}
R \equiv \frac{I_{(2)}}{I_{(1)}},
\end{equation}
and on the nonlinearity $m$. If $R \approx 1$ and $m$ is not sufficiently high, the resulting NISE image contains a ``twin'' image originating from the second brightest speckle. To evaluate how often this unfavourable case occurs, we compute the probability that $R$ is smaller than a threshold $r$.

We assume an exponential probability distribution of speckle-grain intensities in an uncorrelated region~\cite{supp:Goodman2020}:
\begin{equation}
p_I(I) = \frac{1}{\langle I\rangle}\exp\!\left(-\frac{I}{\langle I\rangle}\right),
\qquad I\ge 0,
\label{eq:exp_pdf_M}
\end{equation}
where $\langle I\rangle$ is the average intensity over the illuminated area. The observation area can be divided into $M$ independent speckle grains,
\begin{equation}
M \simeq \frac{A_{\mathrm{ROI}}}{A_c},
\end{equation}
with $A_{\mathrm{ROI}}$ the illuminated area and $A_c$ the mean speckle-grain area.

Under these assumptions (see further sections for the deteailed derivation), the cumulative distribution of the ratio $R$ between the second-brightest and the brightest speckle intensities is
\begin{equation}
P\{R<r\} \;=\; \frac{\Gamma(M+1)\,\Gamma\!\left(1+\tfrac{1}{r}\right)}{\Gamma\!\left(M+1+\tfrac{1}{r}\right)} ,
\qquad 0<r<1,
\label{eq:ratio_cdf}
\end{equation}
where $\Gamma(\cdot)$ denotes the gamma function.

In our experiments, we illuminate the sample with approximately $M\approx 200$ speckle grains. For the threshold $r=0.85$ this yields
\begin{equation}
\mathbb{P}(R<0.85)\approx 0.43.
\end{equation}
With a nonlinear order $m=15$, the signal-to-background ratio (SBR), defined as the ratio of the brightest to the second-brightest luminescence contributions,
\begin{equation}
\mathrm{SBR} \;\equiv\; \frac{I_{(1)}^{\,m}}{I_{(2)}^{\,m}} \;=\; \left(\frac{1}{R}\right)^{m},
\end{equation}
exceeds $10$ for $R\le 0.85$:
\begin{equation}
\left(\frac{1}{0.85}\right)^{15} \approx 11.45 \;>\; 10.
\end{equation}

\subsection{Speckle-field statistics, speckle grain size}
\label{sec:spk_stat_1}
When coherent light propagates through a complex scattering medium, the intensity on an observation plane forms a granular (\emph{speckle}) pattern, a phenomenon well documented in optics \cite{supp:Goodman1975, supp:Goodman1976, supp:Goodman2020}. Although wave propagation is strictly deterministic for a fixed microstructure, the microscopic arrangement, e.g., positions/orientations of individual scatterers, is unknown and uncontrolled; hence, we model the field statistically and take ensemble averages over realizations of the medium.\\
In the fully developed regime (circular complex Gaussian field), pointwise intensities are exponential and correlations obey the Siegert relation, which permits treating a large region of interest (ROI) as $M$ effectively independent speckle grains. This, in turn, enables an order-statistics analysis of the brightest and second-brightest grains, as well as the distribution of their ratio.

Let $I(\mathbf r)$ denote the intensity at transverse coordinate $\mathbf r$. In a fully developed speckle, the pointwise light intensity has an exponential distribution:
\begin{equation}
p_I(I)\;=\;\frac{1}{\langle I\rangle}\,e^{-I/\langle I\rangle}. 
\qquad I\ge 0,
\label{eq:exp_pdf}
\end{equation}

The normalized intensity \emph{autocovariance} obeys the Siegert relation
\begin{equation}
C_I(\boldsymbol\rho)
\;:=\;
\frac{\big\langle \delta I(\mathbf r)\,\delta I(\mathbf r+\boldsymbol\rho)\big\rangle}{\langle I\rangle^2}
\;=\;
\bigl|\mu(\boldsymbol\rho)\bigr|^2,
\label{eq:siegert}
\end{equation}
where $\delta I=I-\langle I\rangle$ and $\mu(\boldsymbol\rho)$ is the complex degree of coherence (field correlation). Consequently, the normalized intensity \emph{autocorrelation} is
\begin{equation}
\Gamma_I(\boldsymbol\rho)
\;:=\;
\frac{\big\langle I(\mathbf r)\,I(\mathbf r+\boldsymbol\rho)\big\rangle}{\langle I\rangle^2}
\;=\;
1+\bigl|\mu(\boldsymbol\rho)\bigr|^2.
\label{eq:int_autocorr}
\end{equation}
After subtracting the unit pedestal, the lobe $\bigl|\mu(\boldsymbol\rho)\bigr|^2$ is the ensemble-average speckle-grain profile in the precise statistical sense: it quantifies how intensities at two points co-vary with separation, and its integral defines the speckle correlation area,
\begin{equation}
A_c \;=\; \iint_{\mathbb R^2} \bigl|(\boldsymbol\rho)\bigr|^2\,d^2\boldsymbol\rho.
\label{eq:Ac_def}
\end{equation}
\noindent\textbf{Key point 1:} \emph{For a region of interest (ROI) of area $A_{\mathrm{ROI}}$ whose linear size greatly exceeds the speckle correlation length, the number of \emph{effectively independent} speckle samples (degrees of freedom) is well approximated by}
\begin{equation}
M \;\approx\; \frac{A_{\mathrm{ROI}}}{A_c}.
\label{eq:Meff}
\end{equation}
This interpretation---that $A_c$ sets the smallest independent ``cell'' and that the density of such cells controls statistics—appears explicitly in the ultrasound and optics literature;
\cite{supp:Wagner1988, supp:RamirezSanJuan2014}.

\subsection{The brightest speckle grain intensity statistics}\label{sec:spk_1st}
Assume the ROI contains $M$ effectively independent speckle grains, ordered by intensities $I_1>I_2>\dots>I_M$. Each speckle grain intensity has a negative exponential distribution [\eqref{eq:exp_pdf}].

Let's define normalized intensity:
\begin{equation}
X_i=I_i/\langle I\rangle
\end{equation}
So $X_i$'s are intensities measured in the units of $\langle I\rangle$. Then each $X_i$ independently of other speckle grains also has an exponential distribution:

\begin{equation}
    p_{X_i}(x)=e^{-x}=f(x),\qquad P_{X_i}(x)=\Pr\{X_i<x\}=F(x), \qquad x\ge 0,
\end{equation}
where \( p_{X_i}(x) \) is the probability density function (PDF) of \( X_i \), and \( P_{X_i}(x) \) is the cumulative distribution function (CDF), denoting the probability that \( X_i \) is less than a particular value \( x \).\\

When we describe an ordered set of speckle grains as a whole, the brightest grain normalized intensity $X_1$ PDF and CDF are:
\begin{align}
P_{X_1}(x)
&= \Pr\{X_1\le x\}
 = \bigl(F(x)\bigr)^M
 = \bigl(1-e^{-x}\bigr)^M, \qquad x\ge 0,\\[4pt]
p_{X_1}(x)
&= \frac{d}{dx}P_{X_1}(x)
 = M\,e^{-x}\,\bigl(1-e^{-x}\bigr)^{M-1}, \qquad x\ge 0.
\end{align}
For the maximum to be $< x$, all $M$ samples must be $< x$, giving $(F(x))^M$ by independence; differentiating it yields the density.

The expected intensity of the brightest speckle $\mathbb{E}\!\left[I_1\right]$ can be calculated by direct algebraic manipulations:
\begin{equation}
\begin{aligned}
\mathbb{E}\!\left[X_1\right]
&=\int_0^\infty \bigl(1-P_{X_1}(x)\bigr)\,dx
=\int_0^\infty \Bigl[1-(1-e^{-x})^{M}\Bigr]\,dx \\
&\overset{t=\,1-e^{-x}}{=}\int_{0}^{1} \frac{1-t^{M}}{1-t}\,dt
\qquad\bigl(e^{-x}dx =dt\Rightarrow dx=\tfrac{dt}{1-t}\bigr)\\
&=\int_{0}^{1}\sum_{m=0}^{M-1} t^{m}\,dt
=\sum_{m=0}^{M-1}\frac{1}{m+1}
=\sum_{m=1}^{M}\frac{1}{m}
=H_{M}.
\end{aligned}
\end{equation}

\begin{equation}
\mathbb{E}\!\left[I_{1}\right] \;=\; \langle I\rangle\sum_{m=1}^{M}\frac{1}{m}
\;=\;\langle I\rangle H_M \;\sim\; \langle I\rangle \left(\ln M + \gamma \right)
\label{eq:avg_brightest}
\end{equation}
where $H_M$ is the $M$th harmonic number and $\gamma$ is the Euler–Mascheroni constant.\\

\noindent\textbf{Key point 1:} \emph{The expected brightest speckle intensity when compared to average intensity grows  \textbf{logarithmically} with the number of effectively independent grains [\eqref{eq:avg_brightest}].}

\subsection{The brightest and the second-brightest intensity statistics.}\label{sec:spk_1st_2nd}

In the same setting, treating the $M$ speckle intensities as independent exponentially distributed, $X_1$ and $X_2$ denote the brightest and the second-brightest speckle grains' normalized intensities. Their joint PDF is

\begin{equation}
p_{X_1,X_2}(x_1,x_2)
= M(M-1)\,e^{-(x_1+x_2)}\bigl(1-e^{-x_2}\bigr)^{M-2},
\qquad 0\le x_2\le x_1<\infty.
\label{eq:joint_top_two}
\end{equation}
One sample falls at $x_2$ (density $f(x_2)$) to be the second maximum, one at $x_1>x_2$ (density $f(x_1)$) to be the maximum, and the remaining $M-2$ samples lie below $x_2$ with probability $F(x_2)^{M-2}$. There are $M(M-1)$ assignments of these roles.

\subsection{Intensity ratio R = X2/X1.}\label{sec:spk_ratio}
In our imaging setting we care how dominant the brightest speckle is relative to the second brightest speckle; a natural dimensionless summary is the ratio \(R\in(0,1)\).
With the change of variables \((x_1,x_2)=(x_1,rx_1)\) (Jacobian \(=x_1\)) applied to \eqref{eq:joint_top_two}, the joint density of \((R,X_1)\) is
\begin{equation}
p_{R,X_1}(r,x_1)
= M(M-1)\,x_1\,e^{-(1+r)x_1}\bigl(1-e^{-rx_1}\bigr)^{M-2},
\qquad x_1>0,\;0<r<1.
\end{equation}
Which needs to be integrated over $x_1$ to recover $f_R(r)$.\\
A more direct route to the CDF uses conditionals:
\begin{equation}
\Pr\!\big\{X_2\le x_2 \,\big|\, X_1=x_1\big\}
= \left(\frac{F(x_2)}{F(x_1)}\right)^{\!M-1}, \quad 0\le x_2\le x_1,
\qquad
p_{X_1}(x_1)=M F(x_1)^{M-1} f(x_1).
\end{equation}
Hence
\begin{align}
P_R(r)
&= \Pr\!\left\{R < r\right\}
= \int_0^\infty \left(\frac{F(rx)}{F(x)}\right)^{\!M-1} p_{X_1}(x)\,dx \nonumber\\
&= M \int_0^\infty \bigl(F(rx)\bigr)^{M-1} f(x)\,dx
= M \int_0^\infty \bigl(1-e^{-rx}\bigr)^{M-1} e^{-x}\,dx. \label{eq:FR_int}
\end{align}
With the substitution \(t=e^{-rx}\) (so \(dt=-r\,t\,dx\) and \(e^{-x}=t^{1/r}\)), \eqref{eq:FR_int} becomes the Beta integral:
\begin{equation}
P_R(r)
= \frac{M}{r}\int_0^1 (1-t)^{M-1}\,t^{\frac{1}{r}-1}\,dt
= \frac{M}{r}\,\mathrm{B}\!\left(\frac{1}{r},\,M\right)
= \frac{M}{r}\,\frac{\Gamma(1/r)\,\Gamma(M)}{\Gamma(M+1/r)}.
\label{eq:FR_closed}
\end{equation}
\begin{equation}
P_R(r)= \frac{\Gamma(1/r+1)\,\Gamma(M+1)}{\Gamma(M+1/r)},
\qquad r>0, \;(\,M>1\,). \label{eq:R_CDF}
\end{equation}

\begin{figure}[ht!]
    \centering
    \includegraphics[width=0.75\linewidth]{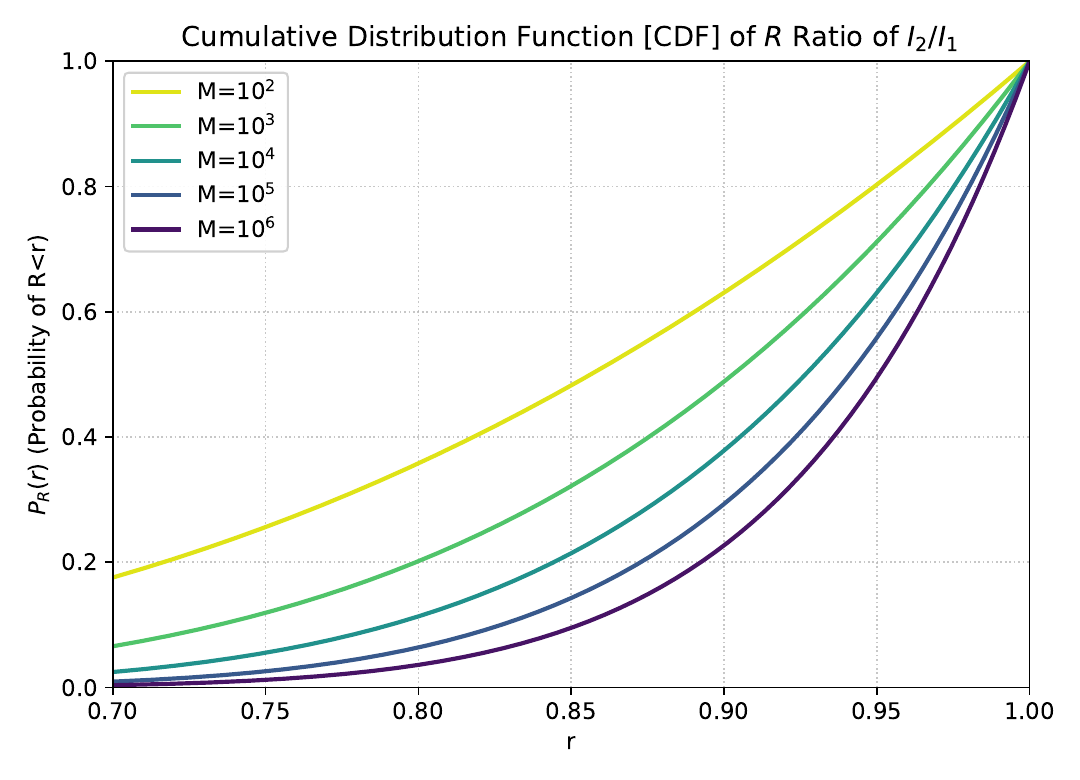}
    \caption{\textbf{CDF of the ratio $R$.} For large $M$, the cumulative distribution function decreases slowly; even for $10^{6}$ independent speckles, $\text{Pr}\{R<0.9\}\approx 23\%$.
}
    \label{fig:R_CDF_GRAPH}
\end{figure}

\noindent\textbf{Key point 2:} The dependency given by \eqref{eq:R_CDF} makes it probable that even if we have an enormous amount of individual speckle grains on the order of $(10^6)$ probability that the brightest will be dominating enough is frequent.

\section{Speckle grain size inside scattering media}
\label{sec:spk_spatial_extent}
In this section, we investigate the behavior of an illuminating beam as it propagates through a scattering medium. Section~\ref{sec:RTE} introduces the wide-angle energy transfer function and its solutions, providing the tools to describe the statistical evolution of light in the scatterer. Then in Section \ref{sec:IROI} we characterize the average intensity of illumination at depth $z$.  In Section~\ref{sec:k_increase} we examine the speckle size in the multiple-scattering regime, which then enables a concise characterization of NISE effective illumination in Section~\ref{sec:nise_illum}.

\subsection{Radiative Transfer Equation, wide-angle formulation}
\label{sec:RTE}
To describe the propagation of the illuminating beam, we model a statistically homogeneous random
scatterer using a \emph{wide-angle transverse-moment} formulation. Rather than invoking a small-angle
(paraxial) approximation, we start from the radiative transfer equation (RTE) and expand its angular
dependence in Legendre moments $P_\ell(\cos\theta)$ (the $P_N$ method), truncated at $\ell\le 2$.
The medium is characterized by:

\begin{description}
  \item[$\mu_s$] \textbf{Scattering coefficient} ($\mathrm{length}^{-1}$): the mean number of scattering
  events per unit path length. The microscopic mean free path is $\ell_s = 1/\mu_s$.

  \item[$g_1$] \textbf{Anisotropy factor} (first Legendre moment of the phase function):
  \begin{equation}
  g_1 \;=\; \langle \cos\theta \rangle \in [-1,1],
  \end{equation}
  with $g_1\approx 0$ for isotropic, $g_1>0$ for forward-peaked, and $g_1<0$ for backward-peaked scattering.

  \item[$g_2$] \textbf{Second Legendre moment} of the phase function:
  \begin{equation}
  g_2 \;=\; \big\langle P_2(\cos\theta)\big\rangle \;=\; \tfrac12\!\left(3\,\langle\cos^2\theta\rangle-1\right).
  \end{equation}
  For the Henyey–Greenstein phase function, one has the identity $g_2=g_1^2$, which is applicable in biological scatterers.
\end{description}

Depth is denoted by $z$ (measured from the entrance plane), with the mean optical axis along $+z$.
Let $r\equiv r_x$ be one transverse coordinate and let
\begin{equation}
\hat{\Omega}=(u_x,u_y,\mu)\in\mathbb S^2,\qquad
\mu=\cos\theta,\qquad u_x=\sin\theta\cos\phi,\quad u_y=\sin\theta\sin\phi,
\end{equation}
so $u_x^2+u_y^2+\mu^2=1$. In what follows we write $u\equiv u_x$ (one transverse
direction cosine); results are identical for $y$ in a laterally isotropic medium.
The first two angular moments
relax exponentially with depth $z$ as
\begin{align}
\beta &:= \mu_s(1-g_1), & \frac{d}{dz}\,\langle u\rangle &= -\beta\langle u\rangle,
\quad\Rightarrow\quad  \langle u\rangle(z) = \langle u\rangle_0\,e^{-\beta z}, \\
\gamma &:= \mu_s(1-g_2), &
\frac{d}{dz}\,\langle u^2\rangle &= -\gamma\!\left(\langle u^2\rangle-\tfrac13\right),
\quad\Rightarrow\quad
\langle u^2\rangle(z)=\tfrac13+\big(\langle u^2\rangle_0-\tfrac13\big)e^{-\gamma z}.
\end{align}
The length $\ell^* := 1/\beta = 1/[\mu_s(1-g_1)]$ is the \emph{transport} (reduced) mean free path.
More generally, the $\ell$-th Legendre moment obeys
\begin{equation}
\big\langle P_\ell(\cos\theta)\big\rangle(z)
= \big\langle P_\ell(\cos\theta)\big\rangle_0\,\exp\!\big[-\mu_s\big(1-g_\ell\big)\,z\big],
\end{equation}
so $\beta$ and $\gamma$ are the exact decay rates of the first and second angular moments, respectively.

Coupled with the streaming identities for any single transverse Cartesian coordinate $r$ and its
conjugate direction component $u$,
\begin{equation}
\frac{d}{dz}\,\langle r\rangle = \langle u\rangle,
\qquad
\frac{d}{dz}\,\langle r^2\rangle = 2\,\langle r\,u\rangle, \qquad \frac{d}{dz}\langle r u\rangle \;=\; \langle u^2 \rangle \;-\; \beta\,\langle r u \rangle,
\end{equation}
with the latter equation derived from collision, these relations yield a closed, linear ODE system for the set
$\{\langle u\rangle,\langle u^2\rangle,\langle r\,u\rangle,\langle r^2\rangle\}$ that remains
\emph{wide-angle} (no paraxial assumption) yet is algebraically tractable. Conceptually, this is the
$P_N$ (spherical-harmonics) formulation truncated at $\ell\le 2$, which is exact for the first and
second angular moments and connects directly to standard RTE treatments in optical transport and
biomedical optics \cite{supp:Goudsmit1940, supp:Duderstadt1979, supp:Markel2004, supp:Liemert2011, supp:Martelli2022}.

\paragraph{Solutions to RTE equations}
The exact solution to this ODE system is given by:
\begin{align}
\quad
&\beta=\mu_s(1-g_1),\quad \gamma=\mu_s(1-g_2),\\
&\langle r\rangle_0,\ \langle u\rangle_0,\ \langle u^2\rangle_0,\ \langle r u\rangle_0,\ \langle r^2\rangle_0\ \text{at }z=0,\\
& \Delta:=\langle u^2\rangle_0-\tfrac{1}{3}, \nonumber\\[4pt]
\langle u\rangle(z) &= \langle u\rangle_0\,e^{-\beta z}, \label{eq:u_sol}\\[4pt]
\langle u^2\rangle(z) &= \tfrac{1}{3} + \Delta\,e^{-\gamma z}, \label{eq:u2_sol}\\[4pt]
\langle r\rangle(z) &= \langle r\rangle_0 + \frac{\langle u\rangle_0}{\beta}\bigl(1-e^{-\beta z}\bigr), \label{eq:r_sol}\\[4pt]
\langle r u\rangle(z) &=
\langle r u\rangle_0\,e^{-\beta z}
+\frac{1}{3\beta}\bigl(1-e^{-\beta z}\bigr)
+\frac{\Delta}{\beta-\gamma}\bigl(e^{-\gamma z}-e^{-\beta z}\bigr), \label{eq:ru_sol}\\[4pt]
\langle r^2\rangle(z) &=
\langle r^2\rangle_0
+\frac{2z}{3\beta}
-\frac{2}{3\beta^2}\bigl(1-e^{-\beta z}\bigr)
+\frac{2\Delta}{\beta-\gamma}\!\left(\frac{1-e^{-\gamma z}}{\gamma}-\frac{1-e^{-\beta z}}{\beta}\right)
+\frac{2\langle r u\rangle_0}{\beta}\bigl(1-e^{-\beta z}\bigr).
\label{eq:r2_sol}
\end{align}

\paragraph{Deep-propagation asymptotics} At depth $z$ greater than $1/\beta$, we can simplify this solutions:
\begin{align}
\langle u\rangle(z)      &= \mathcal{O}\!\left(e^{-\beta z}\right)\;\to\;0,\\
\langle u^2\rangle(z)    &= \tfrac{1}{3} + \mathcal{O}\!\left(e^{-\gamma z}\right)\;\to\;\tfrac{1}{3},\\
\langle r\rangle(z)      &= \langle r\rangle_\infty + \mathcal{O}\!\left(e^{-\beta z}\right)\;\to\;\langle r\rangle_\infty,\\
\langle r u\rangle(z)    &= \tfrac{1}{3\beta} + \mathcal{O}\!\left(e^{-\min\{\beta,\gamma\} z}\right)\;\to\;\tfrac{1}{3\beta},\\
\langle r^2\rangle(z)    &= \frac{2}{3\beta}\,z + \mathcal{O}(1)\;\sim\;\frac{2}{3\beta}\,z\;\xrightarrow[z\to\infty]{}\;\infty\ \text{(linearly)}.
\end{align}

\subsection{Intensity of the area illuminated by speckles}
\label{sec:IROI}
In the small-angle, many-collision limit ($z\gg \ell^*$ with $\ell^*=1/\mu_s'$ and $\mu_s'=\mu_s(1-g_1)$), the transverse intensity profile in a plane at depth $z$ is well approximated by a diffusion kernel (with $z$ playing the role of “time”):
\begin{equation}
I_{\mathrm{diff}}(r,z)
\;\propto\;
\frac{1}{4\pi D z}\,
\exp\!\left(-\frac{r^2}{4Dz}\right),
\qquad
D \;\approx\; \frac{\ell^*}{6}.
\label{eq:Idiff}
\end{equation}
Here $D$ is the effective \emph{transverse} diffusion coefficient in the Fokker–Planck reduction. 
\eqref{eq:Idiff} is the Green’s function of $\partial_z I = D\nabla_{\!\perp}^2 I$ and implies
\begin{equation}
\langle r^2\rangle(z) = 4Dz \;\approx\; \frac{2}{3}\,\ell^* z,
\end{equation}
consistent with the wide-angle moment result; see, e.g., \cite{supp:Martelli2022}.

These results imply that, as light propagates through the scatterer, the diffusive footprint expands with rms radius $\propto\sqrt{\ell^* z}$, so any quantile-defined illuminated area grows linearly with depth:
\begin{equation}\label{eq:A_ROIz}
A_{\mathrm{ROI}}(z)\;\approx\; \pi\ \frac{2}{3}  \ell^* z
\end{equation}
\noindent\textbf{Key point 3:} $A_{\mathrm{ROI}}(z)\propto \ell^* z$.

Consequently, for fixed delivered power (up to attenuation/out-scattering factors), the mean intensity over such an ROI decays inversely with $z$:
\begin{equation}\label{eq:avg_Iz}
\langle I\rangle(z)\;\propto\;\frac{1}{A_{\mathrm{ROI}}(z)}\;\propto\;\frac{1}{z}.
\end{equation}
\noindent\textbf{Key point 4:} $\langle I\rangle(z)\propto z^{-1}$.

\subsection{Speckle size in the multiple-scattering regime}
\label{sec:k_increase}
At depths where many forward-peaked scattering events accumulate, each collision imparts a small random angular deflection, broadening the angular distribution. In the small-angle (Fokker–Planck) limit, a central-limit argument implies an \emph{approximately Gaussian} angular power spectrum. Let $\mathcal S_z(\mathbf k_\perp)$ denote the angular power spectrum at depth $z$, normalized so that $\int \mathcal S_z(\mathbf k_\perp)\,d^2k_\perp=1$, with
\begin{equation}
\mathbf k_\perp = k\,\mathbf u_\perp,\qquad k=\frac{2\pi n}{\lambda}.
\end{equation}
Assuming transverse isotropy and denoting by $s^2(z)$ the per-axis variance of a transverse direction cosine (i.e.\ $s^2(z)=\langle u_x^2\rangle=\langle u_y^2\rangle$), we have the Gaussian approximation
\begin{equation}
\mathcal S_z(\mathbf k_\perp)
\;\approx\;
\frac{1}{2\pi\,(k s(z))^2}\,
\exp\!\Bigg[-\,\frac{\|\mathbf k_\perp\|^2}{2\,(k s(z))^2}\Bigg]. \qquad s^2(z)=\langle u^2\rangle(z),
\label{eq:GaussianAPS}
\end{equation}

By the Van Cittert–Zernike (VCZ) relation for a quasi-homogeneous cross section, the mutual coherence in the transverse plane is (up to a constant) the Fourier transform of $\mathcal S_z$:
\begin{equation}
W(\Delta\mathbf r;z)\ \propto\ \int_{\mathbb R^2} 
e^{\,i\,\mathbf k_\perp\cdot \Delta\mathbf r}\,
\mathcal S_z(\mathbf k_\perp)\,d^2k_\perp.
\label{eq:VCZ}
\end{equation}
Evaluating \eqref{eq:VCZ} for \eqref{eq:GaussianAPS} gives a Gaussian spatial coherence,
\begin{equation}
W(\Delta r;z)\ \propto\ \exp\!\Big[-\,\tfrac12\,\big(k\,s(z)\,\Delta r\big)^2\Big],
\qquad \Delta r=\|\Delta\mathbf r\|.
\label{eq:WGaussian}
\end{equation}
Hence the (normalized) intensity correlation $|W|^2$ is also Gaussian, and the $1/e$ speckle-grain radius is
\begin{equation}
\xi_r(z)\ \approx\ \frac{1}{k\,s(z)}
\;=\; \frac{\lambda}{2\pi n\,s(z)}
\;=\; \frac{\lambda}{2\pi n\,\sqrt{\langle u^2\rangle(z)}}.
\label{eq:xiDefinition}
\end{equation}
As multiple scattering develops, the angular distribution becomes isotropic, so that $s^2(z)\!\to\!1/3$ per transverse axis. Via VCZ, the mutual coherence becomes Gaussian, and the average speckle grain profile is Gaussian. The $1/e$ speckle radius
\begin{equation}
\xi_r(z)\;=\;\frac{1}{k\,s(z)}\;=\;\frac{\lambda}{2\pi n\,s(z)}
\quad\longrightarrow\quad
\xi_r(\infty)\;=\;\frac{\sqrt{3}}{k}\;=\;\frac{\sqrt{3}\,\lambda}{2\pi n},
\end{equation}
i.e. it decreases with depth and saturates at a wavelength-scale value.

\noindent\textbf{Key point 5:} \emph{Speckle grains are (approximately) Gaussian in shape in the multiple-scattering regime.}\\
\textbf{Key point 6:} \emph{Speckle size shrinks with depth and approaches a diffraction-scale limit of order $\lambda/n$.}

\section{The brightest speckle inside a scatterer - effective illumination} \label{sec:nise_illum}
\subsection{Intensity}
As noted in \eqref{eq:Ac_def}, over a sufficiently large observation area $A_{\mathrm{ROI}}$, the speckle intensities can be treated as independent, with an effective count
\begin{equation}
M_{\mathrm{eff}}(z)\;\approx\;\frac{A_{\mathrm{ROI}}(z)}{A_c}.
\end{equation}
In the multiple-scattering (diffusive) regime, the illuminated footprint broadens as $A_{\mathrm{ROI}}(z)\propto \ell^*z$, while deep inside the scatterer the speckle correlation area saturates to a constant set by the diffraction scale, cf.\ \eqref{eq:xiDefinition}, so $A_c\sim \mathrm{const}\times(\lambda/n)^2$. Consequently,
\begin{equation}
M_{\mathrm{eff}}(z)\;\propto\; \ell^*z\,\Big(\frac{n}{\lambda}\Big)^{\!2}.
\end{equation}
Because the mean intensity over the effective illumination area scales as $\langle I\rangle(z)\propto 1/A_{\mathrm{ROI}}(z)\propto z^{-1}$, the expected brightest speckle intensity obeys
\begin{equation}
\mathbb{E}\!\left[I_{(M_{\mathrm{eff}})}(z)\right]\;=\;\langle I\rangle(z)\,H_{M_{\mathrm{eff}}(z)}
\;\sim\; \frac{\ln M_{\mathrm{eff}}(z)}{z}
\;\propto\; \frac{\ln \left( \ell^*z\,\left(\frac{n}{\lambda}\right)^{\!2}\right)}{z}\qquad (z\to\infty),
\end{equation}
up to depth-independent constants.

\noindent\textbf{Why this is instrumental for NISE.}
 The brightest-speckle intensity decays only as $\sim (\ln z)/z$, i.e.\ \emph{algebraically with a mild logarithmic correction}, in stark contrast to ballistic or single-scattered focusing, which suffers an \emph{exponential} decay $\propto e^{-z/\ell^*}$ (on top of attenuation).

\subsection{Sub-diffraction imaging}
In the multiple-scattering regime, speckle grains remain \emph{diffraction-limited} and are well approximated by Gaussian [\eqref{eq:xiDefinition}]. 
Let the local excitation intensity near a speckle maximum be
\begin{equation}
I(r;z)\;\propto\;\exp\!\Big(-\frac{r^2}{2\,\xi_r^2(z)}\Big),
\end{equation}
where $\xi_r(z)$ is the $1/e$ speckle radius at depth $z$. For an $n$-th order nonlinear emitter, the effective excitation profile is
\begin{equation}
I_m(r;z)\;\propto\; \big[I(r;z)\big]^m
\;=\;\exp\!\Big(-\frac{m\,r^2}{2\,\xi_r^2(z)}\Big),
\end{equation}
which is Gaussian with a reduced width
\begin{equation}
\xi_{r,m}(z)\;=\;\frac{\xi_r(z)}{\sqrt{m}},
\qquad
\mathrm{FWHM}_n \;=\; \frac{\mathrm{FWHM}_1}{\sqrt{m}}.
\label{eq:sub_diff_inside}
\end{equation}

\noindent\textbf{Key consequence 1:} Nonlinearity narrows the excitation point-spread function by a factor $1/\sqrt{n}$, delivering sub-diffraction resolution relative to the linear (diffraction-limited) speckle width, \emph{independently of depth} $z$ since $\xi_r(z)$ saturates at a wavelength-scale value in deep media.
Practically, this yields super-resolution with standard scanning with the dominant speckles and without wavefront shaping or heavy computational reconstruction.

\section{Speckle grains in the far field beyond a scattering layer}
\label{sec:spk_far_field}
When the scanned target lies beyond the scattering medium, the resulting speckles retain similar shape and size characteristics, which are important for NISE.\\
The average speckle size at distance $d$ from the scattering layer is \cite{supp:Goodman2020}:
\begin{equation}
    \delta(d)\approx \frac{\lambda d}{D}\;\approx\;\frac{\lambda}{2\,\mathrm{NA}},
\end{equation}
where $D$ is the diameter of the illuminated region at the scattering layer and $\mathrm{NA}\approx D/(2d)$ is the effective numerical aperture. \\
The speckle intensity envelope is bell-shaped and can be approximated by a Gaussian near its maximum; hence, the intensity profile of the brightest speckle can be written as:
\begin{equation}
I(r;d)\propto \exp\!\Big(-\frac{r^2}{2\,\delta^2(d)}\Big).
\end{equation}
In the presence of an $m$-th order nonlinear response, the effective excitation width shrinks by a factor of $\sqrt{m}$, behavior similar to speckle grain inside the scatter [\eqref{eq:sub_diff_inside}]:
\begin{equation}
    \delta_m(d)=\frac{\delta(d)}{\sqrt{m}}=\frac{\lambda}{2\,\mathrm{NA}\,\sqrt{m}}\,.
\end{equation}

\noindent\textbf{Key consequence 2:} Nonlinearity narrows the excitation point-spread function by a factor $1/\sqrt{m}$, delivering sub-diffraction resolution relative to the Abbe limit $\frac{\lambda}{2 \text{NA}}$. Practically, this yields super-resolution using standard scanning with the brightest speckles and without wavefront shaping or computational reconstruction.

\section{Experimental Setup}\label{sec:experiment}

The NISE instrumentation is, at its core, a scanning microscope. The scheme of this setup is shown in Figure \ref{ch5_fig:NISE_scheme}.

\begin{figure}[hp]
\centering\includegraphics[width=0.99\linewidth]{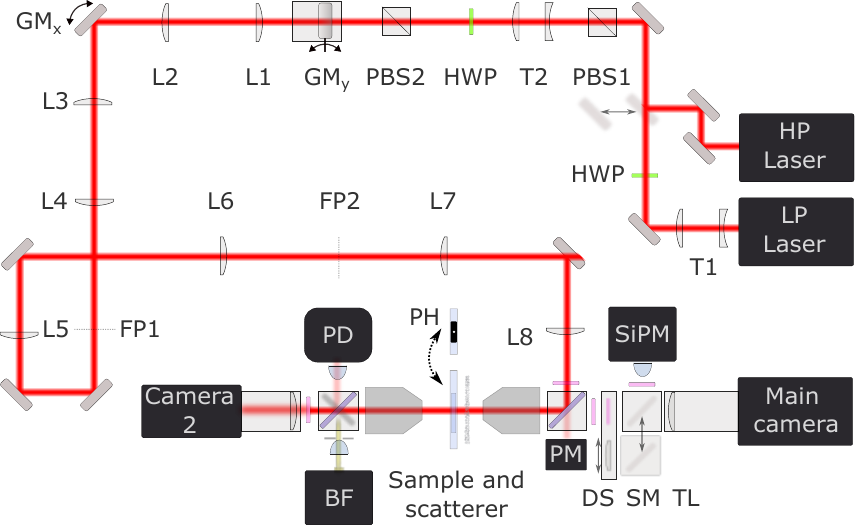}
\captionsetup{font=footnotesize}
\caption{Scheme of the experimental setup for NISE. The scheme represents the real placement of the elements on the optical table (up to scaling). Two CW, 1064nm sources -- Low power laser (3W) and High Power laser (20W) (LP Laser and HP Laser, respectively) are used for excitation of the nonlinear labels. The LP laser beam expands first through the telescope T1 (focal lengths: -30mm, 100mm), and then the polarization is set to be horizontal using a half-wave plate (HWP). HP laser output is unpolarised, and after three steering mirrors, the polarisation is set to horizontal with PBS1. The third mirror is placed on the magnetic base, allowing for switching between the sources. After PBS1, the Galilean telescope T2 further expands the beams (focal lengths: -100 mm, 150mm). Afterwards, a half-wavelength plate in a motorised mount and PBS2 enables precise power control. After PBS2, the beam is elevated using a periscope that ends with a galvo mirror (GM$_y$). This mirror scans along the vertical axis. Next, a 4f system consisting of two lenses with 100 mm focal lengths (L1, L2) images the surface of the GM$_y$ onto the second galvo mirror (GM$_x$). GM$_x$ scans along the horizontal axis. After GM$_x$ scanning along the horizontal axis, the beam traverses another 4f system, consisting of lenses with 100 mm focal lengths (L3, L4), creating a galvo-conjugated plane, FP1. This plane is imaged with a 4f system onto FP2. The first lens in this 4f telescope (L5) has a focal length of 200 mm, and the second lens (L6) has a focal length of 150 mm. FP2 is imaged onto the back focal plane of the objective with a 4f system with L7 lens ($f_{L7} = 100$mm) and L8 lens ($f_{L8} = 200$mm). After L8, the beam goes through a narrowband filter and reflects from a dichroic mirror to the objective. The leakage through the dichroic mirror is collected with a power meter (PM). Behind the objective, the diffuser can be placed completely covering the sample with nonlinear labels. Behind the sample, the excitation is imaged using an additional microscope module, which includes the objective and tube lens (TL2 with a focal length of 100 mm). This imaging process projects the sample plane onto the sensor of Camera2. Before TL2, an absorptive filter reduces the signal, and a dichroic mirror reflects most of the excitation light. This dichroic is also used to introduce the bright-field illumination with an 800 nm LED (BF). The dichroic mirror can be replaced with a silver mirror and direct the light to a photodiode (PD). This configuration is used to take the image of the excitation pattern, by replacing the sample with a pinhole (PH). The Imaging of the sample can occur in two configurations: bright-field imaging with the Main camera, or laser scanning with the Silicon Photon Multiplier (SiPM). For the bright field imaging the 200 mm tube lens creates an image of the sample on the Main camera. For detection with SiPM, a lens with a focal length of 75 mm is placed in detector slider (DS), and the SiPM Mirror (SM) reflects the light into the SiPM detector. Before the detector, a set of spectral filters is followed by a high-NA aspheric lens.}
\label{ch5_fig:NISE_scheme}
\end{figure}

\subsection{Excitation Path}

The excitation path in the NISE supports two complementary sources and preserves well-controlled pupil conjugation through several relays. Two 1064 nm lasers: a high-stability 3 W (CNI MIL-H-1064-3W, EG71048), CW source with TTL gating for fast on–off control (to probe the rise dynamics of avalanching particles, and their luminescence response function $R\{I_{exc}\}$), and a 20 W, CW source (CNI FL-1064-CW-20W) for high-flux operation and imaging through scattering layers (In scheme \ref{ch5_fig:NISE_scheme} LP Laser and HP laser respectively). The 20 W laser is unpolarised, so the beam first passes a polarising beamsplitter (PBS1) to define a linear polarisation, then a half-wave plate and a second PBS for fine power tuning and stable attenuation. The 3W laser is introduced into the same path by removing a mirror from a magnetic mount. The Galilean telescope (T1, plano concave, convex lenses: $f = -30$ mm, $f = 100$ mm) expands the beam so that the lasers have similar beam diameters, and the half-waveplate (Thorlabs: WPH10ME-980) sets the polarization to avoid attenuation at the first PBS.

After power management, a Galilean telescope (T2, plano concave, convex lenses: $f=-100$mm, $f=150$mm) expands either of the beams to fill the objective aperture.

After such preparation, the beam enters the block responsible for scanning. The scanning block is a conjugated two-galvo system (GM$_y$ and GM$_x$, Thorlabs: GVS011/M). One galvo mirror is imaged onto the other with a 4f relay so that the pair behaves as a single pivot at a well-defined plane (FP1, FP2). The 4f relay is built from singlet, 1 inch plano-convex lenses (L1, L2) with focal lengths of 100 mm. With this configuration, the maximum deflection range that preserves a stationary pupil point, i.e., scan angle without lateral beam walk at the conjugate plane, is within approximately $-0.7^{\circ}$ to $+0.7^{\circ}$. In this range, used in the experiment, the excitation beam remains effectively stationary at the pivot, and the paraxial approximation holds.

Downstream of the scanner pair, the beam traverses three additional 4f relays. First relay (plano-convex lenses L3, L4; $f=100$mm) followed by two longer 4f telescopes (plano-convex lenses L5, L6 $f_{L5}$=200 and $f_{L6}$=150, and L7, L8 with achromatic doublet (Thorlabs: AC254-100-B) $f_{L7}$=100, and $f_{L8}$=200mm singlet lens that complete the pupil conjugation to the objective.
The second 4f with 100 mm lenses (L3, L4) ensures symmetry of the two scan axes so that GM$_y$ and GM$_x$ see identical optics.
The third 4f (L5, L6) translates the conjugate scanner plane forward toward the microscope body. The fourth 4f (L7, L8) images the scanner pivot precisely onto the objective’s back focal plane, ensuring that angular deflection at the mirrors produces a pure field tilt at the pupil and a clean lateral shift of the focus in the sample.

Immediately before the beam enters the light-tight microscope body, a narrow notch filter (TECHSPEC 1064 CWL, OD 4.0, 10 nm Bandpass Filter) is placed to suppress any residual wavelengths other than 1064 nm. This also keeps the detection bands free from the stray light.

\subsection{Detection}

In the presented setup, the objective threads into a rigid, custom body 
made of two magnetic cubes and one filter slide. 

Inside the first cube (Thorlabs: DFM1/M), we place a short-pass dichroic with a 900 nm cut-off (Thorlabs: DMS900R) as the main separator, directing the excitation light to the objective (Nikon CFI S Plan Fluor ELWD 20XC). The first magnetic cube holds the objective horizontally through SM1/M25x0.5 adapter (Thorlabs SM1A73). In the next port (opposite to the excitation port) the residual leakage through the dichroic mirror is collected with a power meter (Thorlabs PM16-121). The next port (opposite to the objective) is tightly connected with the detection slider (Thorlabs: CFS1).

For optimal light collection with Silicon Photon Multiplier (SiPM: Thorlabs PDA41) a focusing lens (plano-convex $f = 75$mm) is introduced in the detection slider. Tightly connected to detection slider is the second magnetic cube (Thorlabs: DFM1/M), which enables quick and repeatable reconfiguration between detection branches: either the Main Camera or the SiPM. The mirror inserted inside this cube redirects the sample luminescence onto the SiPM path (non-descanned point detection). A blank 3D-printed insert allows the light to pass through the cube to the wide-field camera path.

The SiPM path begins with a flat mirror  (Thorlabs: DFM1T4). An additional stack of short-pass and notch filters is placed in front of the SiPM, removing any remaining 1064 nm leakage before the detector (in total, in the SiPM detection path, there are four filters: 2X FESH0900, and 2XNF1064-44). Next, a plano-convex lens ($f=25.4$~mm) focuses the luminescence onto the SiPM sensor. For NISE scans we use the SiPM sensor with the highest possible gain ($5\times10^6$) set.   

\subsection{Examination of the speckle pattern}

The speckle fields generated by either a $0.5^{\circ}$ holographic diffuser (Edmund Optics, \#47-988) or a fresh calf brain tissue slice were characterized using an auxiliary microscope module. The module is mounted on a translation rail and can be positioned relative to the sample using a micrometer screw. A Nikon N10X-PF objective collects the transmitted excitation light. Most of the collected light is reflected by a dichroic mirror to minimize back-reflections toward the sample. The residual transmitted light is further attenuated with an absorptive neutral-density filter (Thorlabs: NENIR10A). The speckle pattern can be imaged onto a CMOS camera (Basler: daA3840-45um) using a 100 mm achromatic lens (Thorlabs: AC254-100-B). Together, the objective and lens form a 4f imaging system that relays the speckle pattern from the objective's focal plane plane to the camera sensor.

This imaging configuration is used to verify the memory effect range. For scattering layers investigated in this work, the speckle pattern remains unchanged over the full scanning range employed in the experiments (maximum range of $-0.7^{\circ}$ to $+0.7^{\circ}$). The same imaging system is also used to confirm the absence of a detectable ballistic component transmitted through the holographic diffuser.

To measure the excitation speckle pattern quantitatively, the sample is replaced with a (1~$\mu$m or 2~$\mu$m) pinhole (Edmund Optics, \#56-272 or \#38-536) positioned at the sample plane. The excitation beam is scanned using the microscope galvo mirrors while the light transmitted through the pinhole is recorded with a photodiode (marked on the \ref{ch5_fig:NISE_scheme} as PD). Because the speckle pattern remains invariant during scanning within the memory-effect range, scanning the speckle field across a stationary pinhole is equivalent to translating the pinhole laterally through a stationary speckle pattern.

Accurate positioning of the pinhole relative to the sample plane requires a one-time calibration procedure. First, a reference image of the avalanching particles is acquired without the scattering layer. The auxiliary microscope module is then adjusted such that the particles are in focus in the bright-field inspection arm. The position of the module is subsequently fixed and used as a reference for the calibration. The sample is replaced with the pinhole, and the pinhole surface is brought into focus of camera2 using only the stage on which the pinhole is mounted. The position of the auxiliary microscope remains unchanged throughout this procedure.

Next, the pinhole is translated along the optical axis using a motorized translation stage (Thorlabs MTS25/M-Z8). The axial position corresponding to the smallest measured excitation spot is identified. This procedure determines the offset between the bright-field imaging plane and the plane sampled by the pinhole. In our implementation, this offset is approximately 40~$\mu$m, consistent with the thickness of the pinhole substrate. Because this calibration relies solely on the auxiliary microscope module, it needs to be performed only once.

For speckle measurements, the sample is first brought into focus in the bright-field inspection arm by moving the inspection module. Next, the inspection module's position is fixed and the sample is replaced with the pinhole. Using the same bright-field reference the pinhole is brought to the focus of camera2. Finally, the pinhole is translated by the previously determined axial offset. Speckle patterns are recorded for a series of axial positions around this nominal plane. The positioning uncertainty is primarily limited by the stage backlash, specified to be below 6~$\mu$m. The speckle plane that best reproduces the experimentally observed excitation conditions is selected by comparison with simulated images.

\subsubsection{Image Simulation Procedure}

To simulate NISE images, the experimentally acquired ground-truth image is first resampled to a pixel size of 100 nm using linear interpolation. A Richardson–Lucy deconvolution is then applied to estimate the underlying high-resolution object distribution. Residual background is removed by setting all pixel values below 0.2 to zero after normalizing the image to its maximum intensity.

The measured speckle pattern is resampled to the same 100 nm pixel grid. To account for the nonlinear response of the avalanching microcrystals, the speckle intensity distribution is raised to the tenth power, corresponding to the measured nonlinearity of the avalanching microcrystals. The resulting effective excitation profile is then convolved with the processed ground-truth object to generate the simulated NISE image. The exact processing pipeline can be followed in the accompanying code repository Code~1 (Ref.~\cite{supp:github_Szczypkowski2026}), which was used to prepare the simulations Fig. 1 and analyze the experimental data presented on Fig. 3.

\subsection{Bright-Field}
In the camera branch 200 mm tube lens (Thorlabs TTL-200A) forms the image on a Hamamatsu ORCA-Quest sCMOS sensor. This path provides bright-field preview, and helps to find "best effort" focus. 

To introduce bright-field illumination, we use an 800 nm LED (BF) in a Köhler-like configuration. For that, we use an additional microscope module that also functions as an additional inspection device. The LED active area is imaged onto the back focal plane of the objective with a 30~mm lens. Right after the lens, an aperture is placed to restrict the illuminated area. 
The BF module is mounted on a rail and can freely move along the optical axis. The objective, acting here like a condenser lens, is mounted on an x-y translation mount (Thorlabs: CXY1A).

\section{Avalanching Particles} \label{sec:avalanching_particles}
\begin{figure}[ht!]
  \centering
  \includegraphics[width=\linewidth]{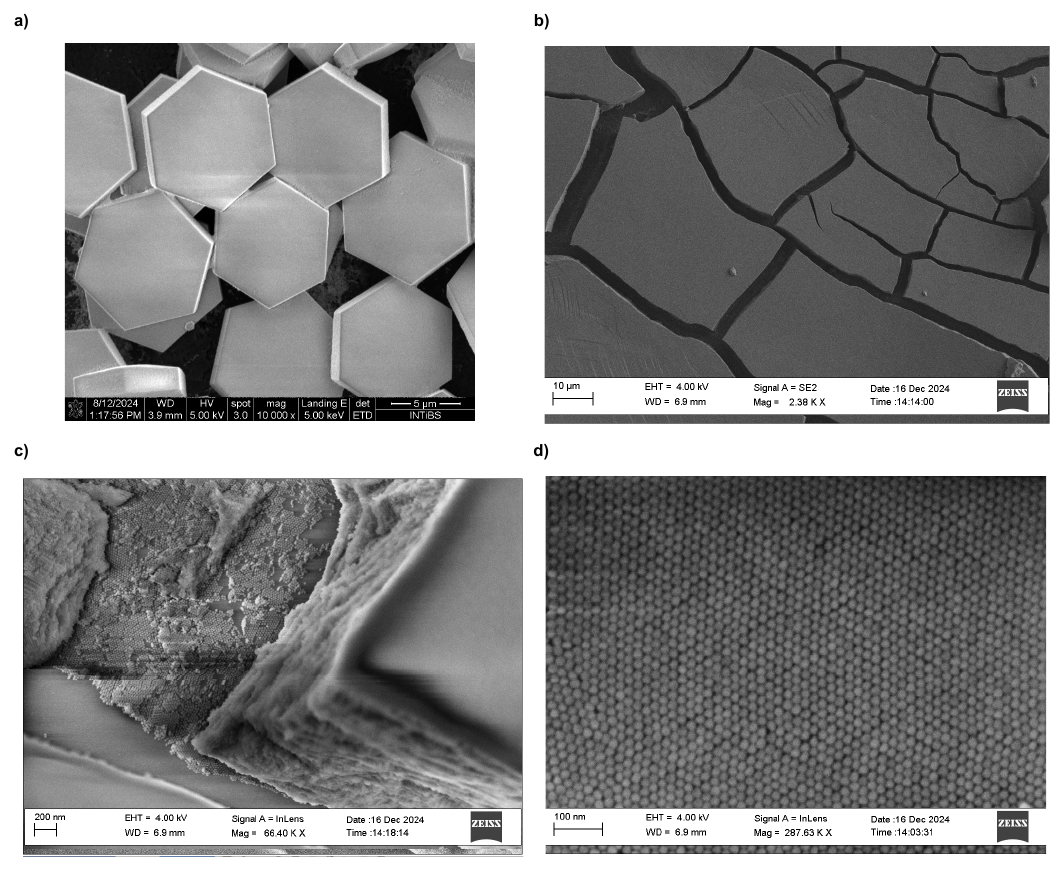}
\caption{Scanning electron microscope images of the avalanching particles. (a) Avalanching microcrystals. (b) Irregular layer of ANPs. (c) Closeup photo of the domain edge of the ANPs sample. (d) The flakes/domain regular surface. Scale bars: (a) $5\,\mu\mathrm{m}$, (b) $10\,\mu\mathrm{m}$, (c) $200\,\mathrm{nm}$, (d) $100\,\mathrm{nm}$.
}
\label{fig:sup3}
\end{figure}
In our experiments, we employed avalanching particles (ANPs) as the active medium exhibiting photon avalanche (PA) behavior. The photon avalanche process arises in heavily lanthanide-doped dielectric materials, where efficient excited-state absorption (ESA) and negligible ground-state absorption (GSA) enable cyclic energy transfer between neighbouring lanthanide ions. This feedback mechanism leads to a steep, nonlinear increase of the PA luminescence intensity with excitation power, typically following a power law with an exponent greater than 5. The emission initiates above a distinct pump threshold and saturates at higher excitation levels, displaying exceptional photostability and narrow spectral features due to the intrinsic properties of Ln$^{3+}$ ions.
Two types of avalanching particles have been synthesized and studied, aiming to demonstrate the NISE features. Firstly, large, hexagonal and flat in shape \ce{NaYF4} photon avalanche microcrystals (AMPs) (Fig. \ref{fig:sup3}a), which were doped with 8\% \ce{Tm^3+} ions (which replace \ce{Y^3+} ions of the host crystal). The second type of particles are photon avalanche nanoparticles (ANPs) of the same composition. These ANPs show good and narrow size distribution with an average diameter of ca. 25~nm (Fig.\ref{fig:sup3}d). While in our case the ANPs form a layer, this type of nanoparticle, after appropriate biofunctionalization, can be potentially used to target and label particular cells or organelles. AMPs and ANPs were characterized using scanning electron microscopy (Fig. \ref{fig:sup3}).

\subsection{Sample Preparation}
\label{sec:methods_sample_preparation}
The \ce{NaLuF4} microparticles doped with 8 mol \% \ce{Tm^3+} ions were prepared using the hydrothermal method based on previously reported synthesis~\cite{supp:Lin2017}. 3 mmol of citric acid (2 M, 1.5 mL), 5 mmol of NaOH (4 M, 1.25 mL), and 10 mL of deionized water were mixed and stirred for 10 min. Then (1 mmol, 1.29 ml) \ce{Ln(NO3)3} (0.92 mmol of \ce{Lu(NO3)3}, 0.08 mmol of \ce{Tm(NO3)3)} were added to the mixture and then stirred for 30 min to form the Ln-Cit$^{3-}$ complex. Subsequently, 16 mL of aqueous solution containing 9 mmol of NaF (1 M, 9 mL) and 7 mL of deionized water were added to form a colloidal suspension and kept stirring for another 30 min. The mixture was transferred to a 50-mL Teflon vessel and heated to 180~$^\circ$C for 12 h. After being cooled to room temperature, the reaction product was isolated by centrifugation and washed with ethanol and water. The sample that was used for imaging was prepared by a drop-casting method. Scanning electron microscope image of the microcrystals are presented in Figure \ref{fig:sup3} a.

The hexagonal-\ce{NaYF4}: 8\%\ce{Tm^3+} nanoparticles were synthesized through a thermal decomposition reaction of lanthanide oleates. To create the precursor, lanthanide acetates [\ce{(CH3COO)3Y} and \ce{(CH3COO)3Tm}] were prepared by mixing stoichiometric amounts of \ce{Y2O3} and \ce{Tm2O3} with a 50\% solution of acetic acid. This mixture was stirred and heated until a clear solution formed. The final precursor was obtained by evaporating the solvents under reduced pressure and then drying it at 140 $^\circ$C for 12 hours.

For the nanoparticle synthesis, 2.5 mmol of the acetates [(CH\(_3\)COO)\(_3\)Y and \ce{(CH3COO)3Tm}] were added to a flask with 15 ml of oleic acid and 38 ml of octadecene. The mixture was stirred and heated to 140 $^\circ$C under vacuum for 30 minutes to create an oleate complex and eliminate any oxygen and residual water. The temperature was then reduced to 50 $^\circ$C, and 10 mmol of ammonium fluoride (\ce{NH4F}) and 6.25 mmol of sodium hydroxide (NaOH) dissolved in 20 ml of methanol were added to the reaction flask. This mixture was stirred for 30 minutes at 70 $^\circ$C. Afterwards, the temperature was raised, and the methanol was evaporated. Once the methanol was removed, the solution was heated to 300 $^\circ$C in a nitrogen atmosphere and maintained at this temperature for one hour. The mixture was then allowed to cool to room temperature.

The nanoparticles were precipitated using ethanol, centrifuged at 10000 rpm for 10 minutes,
and washed with hexane and ethanol. Finally, the nanoparticles were dispersed in chloroform,
resulting in a stable colloidal solution without any aggregation. The sample that was used for
imaging was prepared by a drop-casting method. Scanning electron microscope images of dried nanoparticles are
presented in Fig. \ref{fig:sup3} b,c,d.

\section{Optical characterization of the nonlinear luminescence of avalanching particles} \label{sec:luminescence_of_avalanching}
\begin{figure}[ht!]
  \centering
  \includegraphics[width=0.95\linewidth]{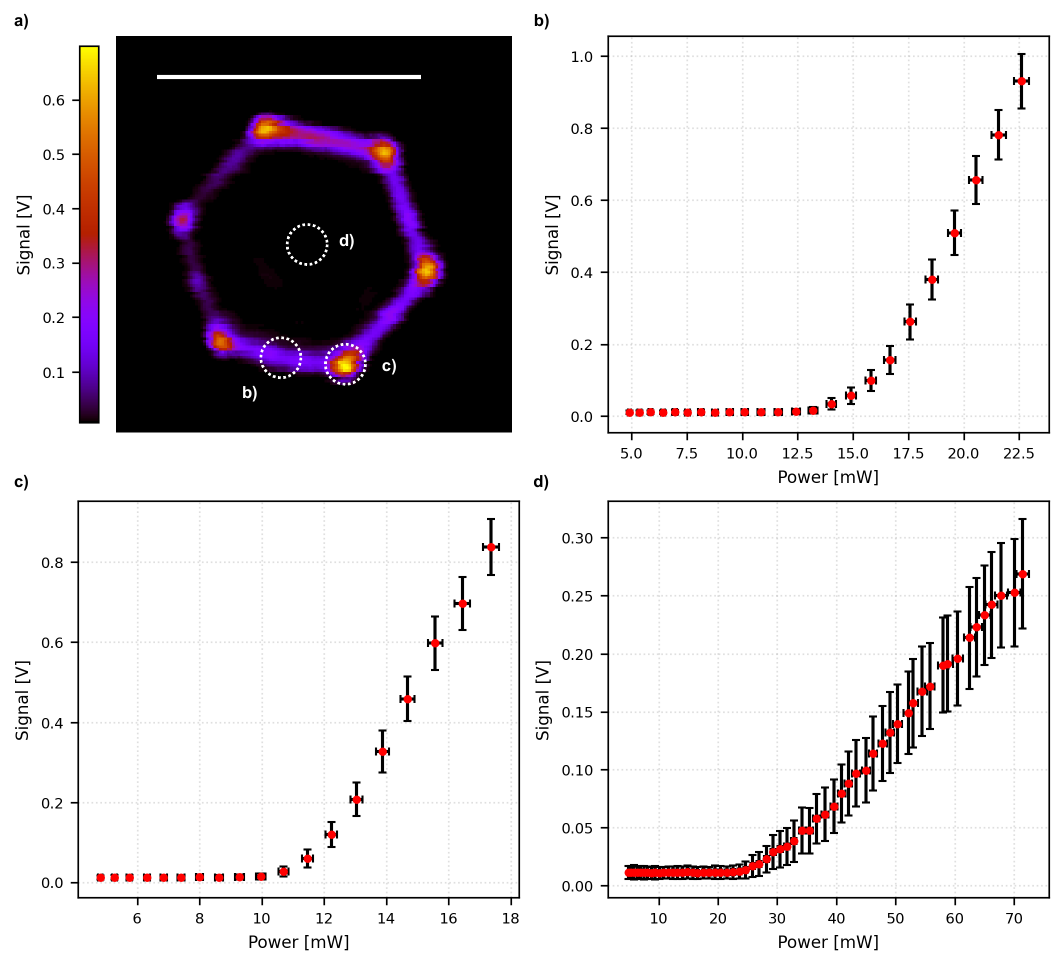}
\caption{Luminescence response of avalanching microcrystals depending on the position of excitation. a) Scanning microscope image, with marked places where the luminescence was measured (edge, corner, face of the avalanching microcrystal). b) Luminescence response at the edge of the microcrystal. c) Luminescence response at the hexagon corner. d) Luminescence response at the face (middle) of the microcrystal. Scale bar in a) 10 $\mu$m. In b),c),d) the y-axis is the measured SiPM detector signal in volts, and the x-axis is the laser power after a 20X NA=0.5 objective.}
\label{fig:sup2}
\end{figure}

\begin{figure}[ht!]
  \centering
  \includegraphics[width=\linewidth]{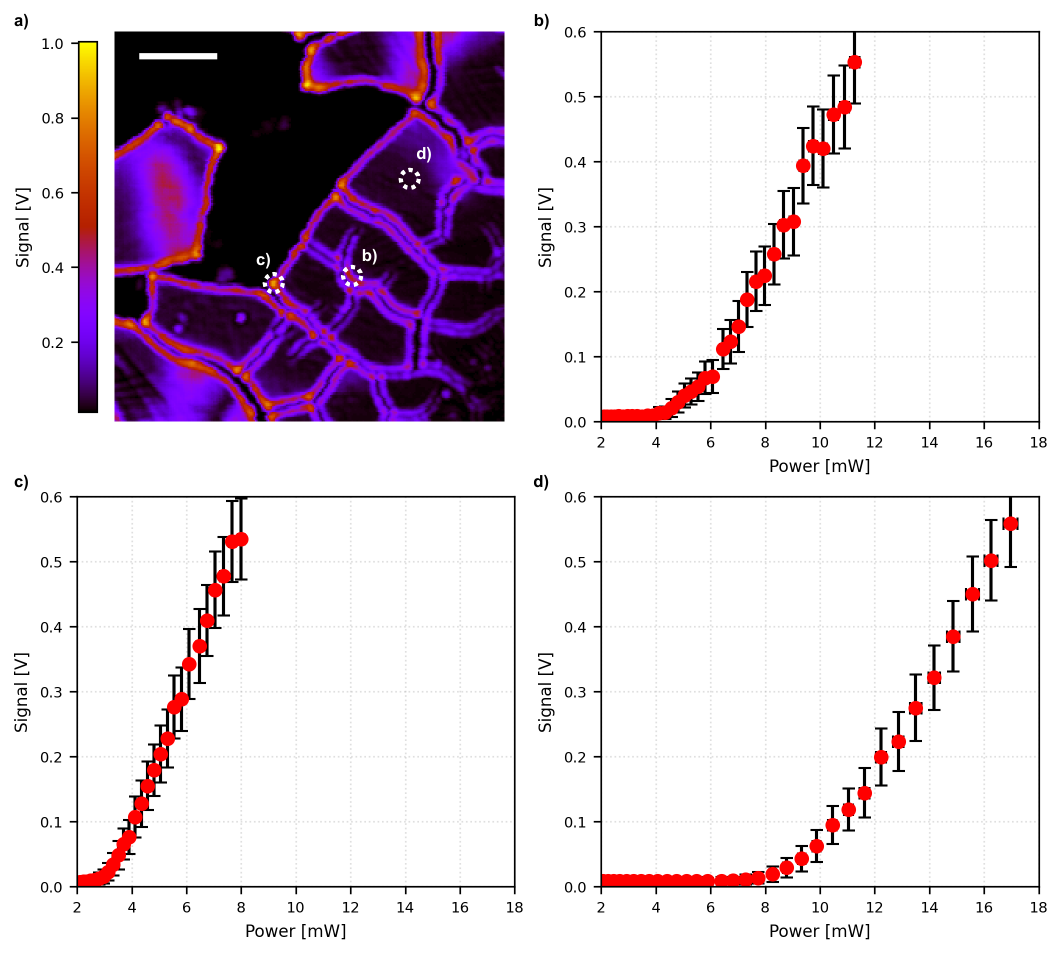}
\caption{Luminescence response of the ANPs layer depending on the position of excitation. a) Scanning microscope image, with marked places where the luminescence was measured (edge, corner, face of the quasicrystalline flakes). b) Luminescence response at the edge of the flake. c) Luminescence response at the corner of the flake/domain. d) Luminescence response at the face of the domain. Scale bar in a) $10\, \mu m$. In b),c),d) The y-axis is the measured SiPM detector signal in Volts, and the x-axis is the laser power after a 20X NA=0.5 objective.
}
\label{fig:sup1}
\end{figure}

The nonlinear optical properties of the avalanching particles were investigated by analyzing the spatially dependent luminescence response under focused laser excitation. The photon avalanche emission exhibits a highly nonlinear dependence on the local excitation intensity, and its magnitude varies with the excitation position within individual microcrystals and domains of the ANPs layer.
In the case of single avalanching microcrystals (Fig. \ref{fig:sup2}), distinct regions such as the crystal edge, vertex, and central face were examined, measuring the luminescence response with respect to the excitation power. First, the image of the sample was taken by scanning the focused beam, with a 20X objective with NA=0.5, a step size of 200 nm, and a 5 ms dwell time. Next, the galvo mirrors were switched off from the power supply to reduce small beam displacements due to electronic noise. The microscope stage was moved to position the excitation beam in the regions of interest (edges, vertices, or faces). With brightfield preview, we confirmed that the luminescence originated from the appropriate regions of interest. To measure the luminescence intensity profiles, the luminescence was directed onto a Silicon Photon Multiplier (SiPM), which output a voltage proportional to the luminescence intensity. For every data point in the plots Fig. \ref{fig:sup2} b, c, d the single measurement took 1 s after the CW laser was enabled with a TTL signal. The DAQ card collected 10000 voltage samples from the SiPM, and then the mean voltage signal was calculated, with errors being the standard deviation of the voltage signal. Next, the excitation power was increased by rotating a half-wave plate placed before the polarising beamsplitter cube. The excitation power was measured with a power meter measuring the laser leakage through the dichroic mirror of the microscope. The measured power was converted into power at the sample with a previously measured conversion factor. The luminescence intensity profiles revealed that the lowest avalanche threshold occurs near structural discontinuities (edges and vertices), compared to the flat crystal faces, with the mechanism of such behavior being under investigation. 
A similar spatial dependence was observed in the ANPs layer (Fig. \ref{fig:sup1}). The emission was probed at the edges, corners, and central regions of the quasi-crystalline flakes. Lower PA thresholds were again detected near the boundaries and corners of the domains.

We find that the optimal power required for imaging in our setup is around 8 mW at the sample with nanoparticles, and around 20 mW at the sample with microparticles (ground truth imaging). While imaging with speckles, we use around 500 mW to image nanoparticles through the calf brain and around 900 mW to image the microparticles through the optical diffuser.

\clearpage

\endgroup

\end{document}